\documentclass[reprint,onecolumn,secnumarabic,amssymb, nobibnotes, aps, prfluids]{revtex4-2}

\usepackage{amsmath}
\usepackage{graphicx}
\usepackage{psfrag}
\usepackage{amssymb}
\usepackage{color}
\usepackage[]{natbib}

\usepackage{epstopdf, epsfig}
\usepackage{amsmath}
\usepackage{subcaption}
\usepackage{xcolor}
\usepackage{bbm}

\newcommand{\andre}[1]{{{#1}}}
\newcommand{\modif}[1]{#1}

\begin{document}

\title{Structure interactions in a reduced-order model for wall-bounded turbulence}%

\begin{abstract}
New reduced-order models (ROMs) are derived for sinusoidal shear flow (also known as Waleffe flow) and plane Couette flow in small periodic domains. A first derivation for Waleffe flow exploits Fourier modes that form a natural orthonormal basis for the problem. A ROM for such basis is obtained by a Galerkin projection of the Navier-Stokes equation. A large basis was reduced to 12 modes that contribute significantly in maintaining chaotic, turbulent dynamics. A key difference from earlier ROMs is the inclusion of two roll-streak structures, with spanwise wavelengths equal to $L_z$ and $L_z/2$, where $L_z$ is the spanwise length of the computational box. The resulting system was adapted to Couette flow by rewriting the Galerkin system for the same 12 modes, modified so as to satisfy no-slip conditions on the walls. The resulting dynamical systems lead to turbulence with finite lifetimes, in agreement with earlier ROMs and simulations in small domains. However, the present models display lifetimes that are much longer than in earlier ROMs, with differences of more than an order of magnitude. The Couette-flow model is compared to results of direct numerical simulation (DNS), with statistics displaying fair agreement.  The inclusion of the $L_z$ and $L_z/2$ lengthscales is seen to be a key feature for longer turbulence lifetimes: neglecting any of the roll modes, or their non-linear interaction, leads to drastic reductions of turbulence lifetimes. The present ROMs thus highlight some of the dominant nonlinear interactions that are relevant in maintaining turbulence for long lifetimes.
\end{abstract}

\author{Andr\'e V. G. Cavalieri}%
\email{andre@ita.br}
\affiliation{Divis\~ao de Engenharia Aeroespacial, Instituto Tecnol\'ogico de Aeron\'autica, S\~ao Jos\'e dos Campos, SP, Brazil}
\date{\today}%
\maketitle

\section{Introduction}

Plane Couette and pipe flow are canonical configurations of wall-bounded flows which transition to turbulence in spite of their stability to infinitesimal disturbances~\citep{schmid2001stability}. It has now been established that the transition to turbulence in such flows is related to the amplitude of the disturbance, and the flow may be maintained in the laminar state for high Reynolds numbers in controlled disturbance environments. Plane channel flow, despite its linear instability setting in at Reynolds number 5772~\citep{orszag1971accurate}, often presents turbulent flow at much lower Reynolds numbers, in a behaviour similar to Couette and pipe flow. Reviews of experimental results showing the amplitude dependence of transition in pipe flow are presented by  \citet{eckhardt2007turbulence} and \citet{mullin2011experimental}.

As the laminar solution of these flows is \modif{linearly stable}, a relevant question is related to what maintains the flow in a turbulent state. Numerical simulations have proven to be useful tools to address this question, particularly as the geometry of the aforementioned canonical flows has two homogeneous directions that allow the use of periodic boundary conditions. The truncation of the computational domain to a small region greatly reduces the number of degrees of freedom of the problem, which simplifies the analysis. For pipe flow the azimuthal discretisation is imposed as between 0 and $2\pi$, but in the axial direction different pipe extents may be imposed. For plane Couette and channel flows, the standard computational domain is a box with lengths $L_x$ and $L_z$ in streamwise and spanwise directions, and the freedom to choose these lengths have motivated a search for minimal flow units for channel \citep{jimenez1991minimal} and Couette flow \citep{hamilton1995regeneration}. These are minimal periodic boxes that are able to maintain turbulence for large times. Analysis of such minimal flow units show that they comprise a single streak of streamwise velocity fluctuations, flanked by nearly streamwise rolls (or streamwise vortices); these stuctures burst intermittently and subsequently reform. Although often designed for low Reynolds number $Re$, minimal flow units may be also used to study turbulence dynamics at higher Re \citep{flores2010hierarchy}, also aiming at the simpler analysis that is possible if a single turbulent structure in the domain dominates the dynamics. Results of such minimal flow units for large Re show phenomena that are similar to what is observed in near-wall units~\cite{hwang2010self}. 

Further insight on the problem of transition and turbulence is possible by simplification of the Navier-Stokes system. Linearisation around a suitable base flow is a common approach. If the laminar solution is taken as the base-flow, one models the evolution of small disturbances, which, despite the linear stability of the aforementioned flows, may result in significant transient growth via the Orr and lift-up mechanisms~\citep{butler1992three,trefethen1993hydrodynamic,reddy1993energy}. Such bounded, transient growth of fluctuations is a key aspect of transition induced by finite-amplitude disturbances, since a sufficiently strong initial perturbation to the flow may be significantly amplified in order to trigger subsequent non-linear effects. When dealing with turbulent flows, it is also possible to linearise the Navier-Stokes system around the mean turbulent profile. An \emph{a priori} justification of the procedure is not straightforward, but it is seen that a similar lift-up mechanism obtained by such analysis leads to agreement with features of turbulent flows~\citep{butler1993optimal,del2006linear,pujals2009note}. \modif{Other analyses are possible if,  instead of taking the mean turbulent profile as a base flow, one considers the linear stability of dominant turbulent structures, such as streaks in minimal turbulent units~\citep{hamilton1995regeneration,schoppa2002coherent}, so as to evaluate mechanisms of streak breakdown}. 

Linear models can thus be quite useful to extract relevant aspects of transitional and turbulent motion, but ultimately there is a need to include \modif{at least some} non-linear effects in order to study how turbulence sustains itself, since fluctuations in the aforementioned linearised models ultimately decay due to the stability of the base flow. Non-linear reduced-order models (ROMs) have thus been derived by truncating the Navier-Stokes system with a small number of spatial modes. The choice of modes may be informed by results of linear analysis, and the ROM so obtained allows a study of the interactions among a finite number of turbulent structures. An early effort was presented by \citet{waleffe1997self}, who considered a wall-bounded flow with free-slip boundary conditions, driven by a streamwise body force. Such configuration, later referred to as Waleffe flow, allows a discretisation using Fourier modes, and a ROM with 8 modes was derived and further reduced to a 4-mode model by assuming some deterministic relations between mode amplitudes. The 4-mode model displays features of streak instability, but does not lead to chaotic motion. Later modelling works were presented by \citet{eckhardt1999transition}, who considered a 19-mode model for a free-slip approximation of Couette flow, and by \citet{moehlis2004low}, who derived a 9-mode model for Waleffe flow. Simulations of both systems reveal a behaviour of transient chaos: the system displays chaotic dynamics for long times, but eventually return to the laminar solution. These are features of a chaotic saddle, with a finite lifetime of chaotic motion, which is seen to increase exponentially with growing Reynolds number.  An extension of the latter model was proposed by \citet{dawes2011turbulent}, who considered larger numbers of Fourier modes in the spanwise direction in the original 8-mode model by \citet{waleffe1997self}, leading to a model with eight partial differential equations. This was seen to considerably change the chaotic saddle, with longer chaotic transients for most initial conditions, triggered by disturbances with lower amplitudes.

Another modelling option is to obtain modes from a direct numerical simulation, usually using proper orthogonal decomposition (POD) \citep{noack2003hierarchy}. This was attempted by~\citet{smith2005low} for a minimal flow unit of Couette flow. The advantage of POD-based models is the use of orthogonal modes that are optimal in representing the kinetic energy in a database; however, such models are known to neglect relevant dynamics and to present numerical instabilities requiring the introduction of additional modelling assumptions, as discussed by \citet{sirisup2004spectral} and \citet{loiseau2019pod}. For instance, in \cite{smith2005low} an eddy-viscosity model is introduced to model neglected POD modes, with a coefficient that is adjusted so as to match dynamics observed in a full simulation.

The observation of finite lifetimes of chaotic motion in reduced models paralleled further research on the transition behaviour of pipe, Couette and channel flow. All these flows are known to have a transition related to finite-amplitude disturbances, with an amplitude threshold proportional to $Re^{-\gamma}$, with $\gamma$ being a positive constant. Experimental results indicate $\gamma=1$ for pipes~\citep{hof2003scaling} and channels~\citep{lemoult2012experimental}, but with different types of disturbance  $\gamma=1.4$ is obtained for the pipe~\citep{mullin2011experimental}.  A careful study of turbulent lifetimes of turbulence induced by application to pipe flow of such small impulsive disturbances, above the critical amplitude, leads to a turbulent pattern, referred to as a puff, which also has a finite lifetime~\citep{hof2006finite}. However, puffs may also split, leading to a larger region of localised turbulent flow, and turbulence becomes self-sustained when the probability of puff splitting becomes higher than the probability of puff decay to the laminar state~\citep{avila2011onset,barkley2016theoretical}. Numerical simulations with sufficiently long domains of pipe flow display such features, but shorter computational domains only display finite turbulence lifetimes, as the domain becomes too small to model the process of puff splitting~\citep{willis2009turbulent}. 

Plane Couette flow is also known to have similar features, with minimal computational domains leading to finite turbulence lifetimes that grow with increasing Reynolds number~\citep{kreilos2014increasing}, similar to the results of ROMs. However, if the domain (or experimental setup) is sufficiently large, turbulence initially develops in oblique patterns, or bands~\citep{bottin1998statistical,duguet2010formation}; as discussed in the reviews of \citet{manneville2015transition} and \citet{tuckerman2020patterns}, these patterns lead to self-sustained turbulence once they start to spread over space. Such behaviour may be captured by reduced-order models truncating the Navier-Stokes system to a small number of modes in the wall-normal direction. ROMs with partial differential equations in the wall-parallel directions were obtained by \citet{lagha2007modeling} for Couette and by \citet{chantry2017universal} for Waleffe flow. Despite their clear interest in obtaining dominant features of transitional and turbulent wall-bounded flows, these models are sets of partial differential equations with numbers of degrees of freedom that remain large, as several streamwise and spanwise wavenumbers are considered in the expansion. Such models include thus a large number of possible non-linear interactions, and the relevant modes and interactions for the dynamics of transition and turbulence are not immediately clear.

The present work revisits reduced-order models for Waleffe and Couette flow in small computational domains such as minimal flow units. It was motivated by the realisation that typical turbulence lifetimes in the 9-mode model by \citet{moehlis2004low} (hereafter referred to as the MFE model) are of about a thousand convective time units, a short duration in comparison with typical time series of direct numerical simulations of minimal flow units that remain turbulent; for instance, \citet{smith2005low} and \citet{nogueira_2021} have analysed minimal flow units of Couette flow with 20000 and 15000 convective time units, respectively, without relaminarisation. Moreover, the chaotic saddle of the MFE model has a fractal behaviour with slight changes of initial conditions leading to either short or long turbulence lifetimes, which is also in contrast with what is found in the simulation of minimal flow units. As discussed above,  Couette flow  \modif{at low Reynolds number} in small computational domains \modif{such as minimal flow units} does not present self-sustained turbulence, but it appears that the MFE model lacks features, or modes, that are relevant in maintaining turbulence for longer lifetimes for randomly chosen initial conditions. This being the case, such features are important components of turbulence dynamics and should be explored in some detail. We anticipate that due to the small computational domains that will be considered, turbulence will not be self-sustained \modif{for the range of parameters considered here}, but the models in the present work, for Waleffe and Couette flows, display turbulence lifetimes that are orders of magnitude higher than the MFE model and thus more compatible with the experience in numerical simulation.

\andre{The reduced-order nature of the model leads to a finite number of non-linear interactions between modes, which become explicit in the model equations. Neglecting some of the non-linearities in the model provides insight on interactions that are relevant to maintain longer turbulence lifetimes. \modif{This is similar in spirit to the restricted non-linear (RNL) models of \citet{farrell2012dynamics} and \citet{thomas2015minimal}}, where the dynamics of streamwise averaged velocities is approximated by neglecting non-linear interactions among wavy disturbances (i.e. streamwise-varying modes), allowing nonetheless to recover the mean velocity profile. On the other hand, some non-linear interactions should of course be retained for accurate turbulence dynamics. \modif{The recent results of  \citet{bae2019nonlinear} indicate}, on the other hand, that some non-linear interactions are crucial, as removal of the projection of the non-linear term onto the leading resolvent forcing mode, which excites rolls, leads to relaminarisation in minimal flow units. In the present model all non-linear interactions appear explicitly in the model equations, and it will be seen that neglect of some of them, either by setting non-linear terms artificially to zero, or by completely neglecting a given mode, leads to significant reduction of turbulence lifetimes.

\modif{The model for Couette flow allows an exploration of the role of non-linear interactions in a configuration that is widely studied as a canonical wall-bounded turbulent flow, with plenty of available results in the literature allowing validation of trends obtained in the reduced-order model with full simulations. The available reduced-order models for Couette flow have limitations in this regard: the model by \citet{eckhardt1999transition} considers free-slip boundary conditions which do not allow comparison with standard simulations or experiments, and the model by \citet{smith2005low} is based on POD modes obtained for a minimal flow unit at Reynolds number 400, and hence cannot be easily applied to other Reynolds numbers or box sizes. The present work provides a ROM for Couette flow with a closed-form basis satisfying no-slip boundary conditions, which may be compared to direct numerical simulations with various computational domains.}}

The remainder of this work is organised as follows. In \S~\ref{sec:derivation} we show how reduced-order models for Waleffe and Couette flow are derived, and results of such models are presented in \S~\ref{sec:results}. As the model results highlight the relevance of interactions between rolls and streaks with different spatial \modif{lengthscales}, this is further investigated in \S~\ref{sec:scaleinteraction}. The paper is completed with conclusions in \S~\ref{sec:conclusions}.


\section{Derivation of reduced-order models}
\label{sec:derivation}

\subsection{Basic definitions}

\begin{figure}
\begin{subfigure}{0.49\textwidth}
\includegraphics[width=1.0\textwidth]{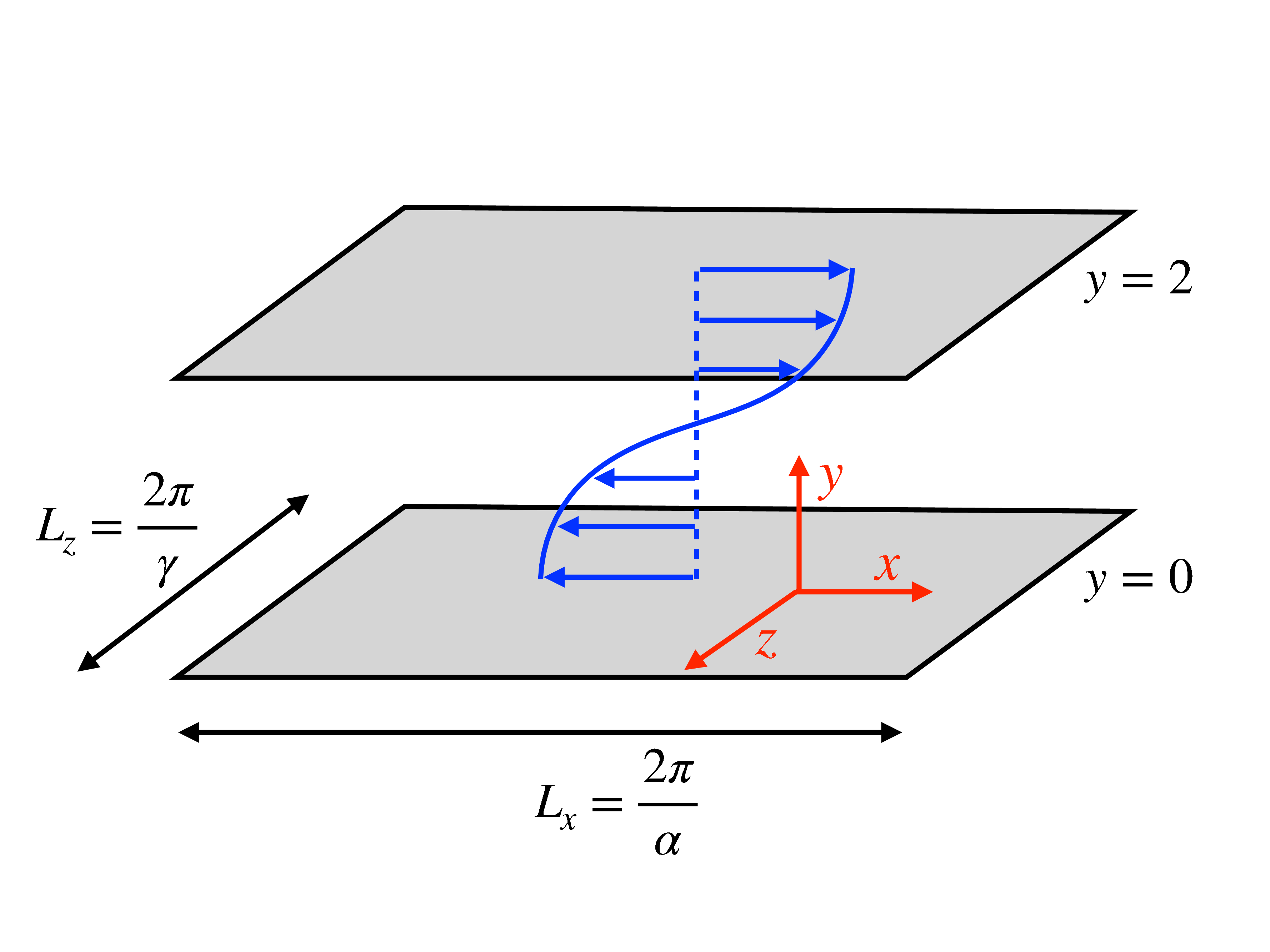}
\caption{Waleffe flow}
\end{subfigure}
\begin{subfigure}{0.49\textwidth}
\includegraphics[width=1.0\textwidth]{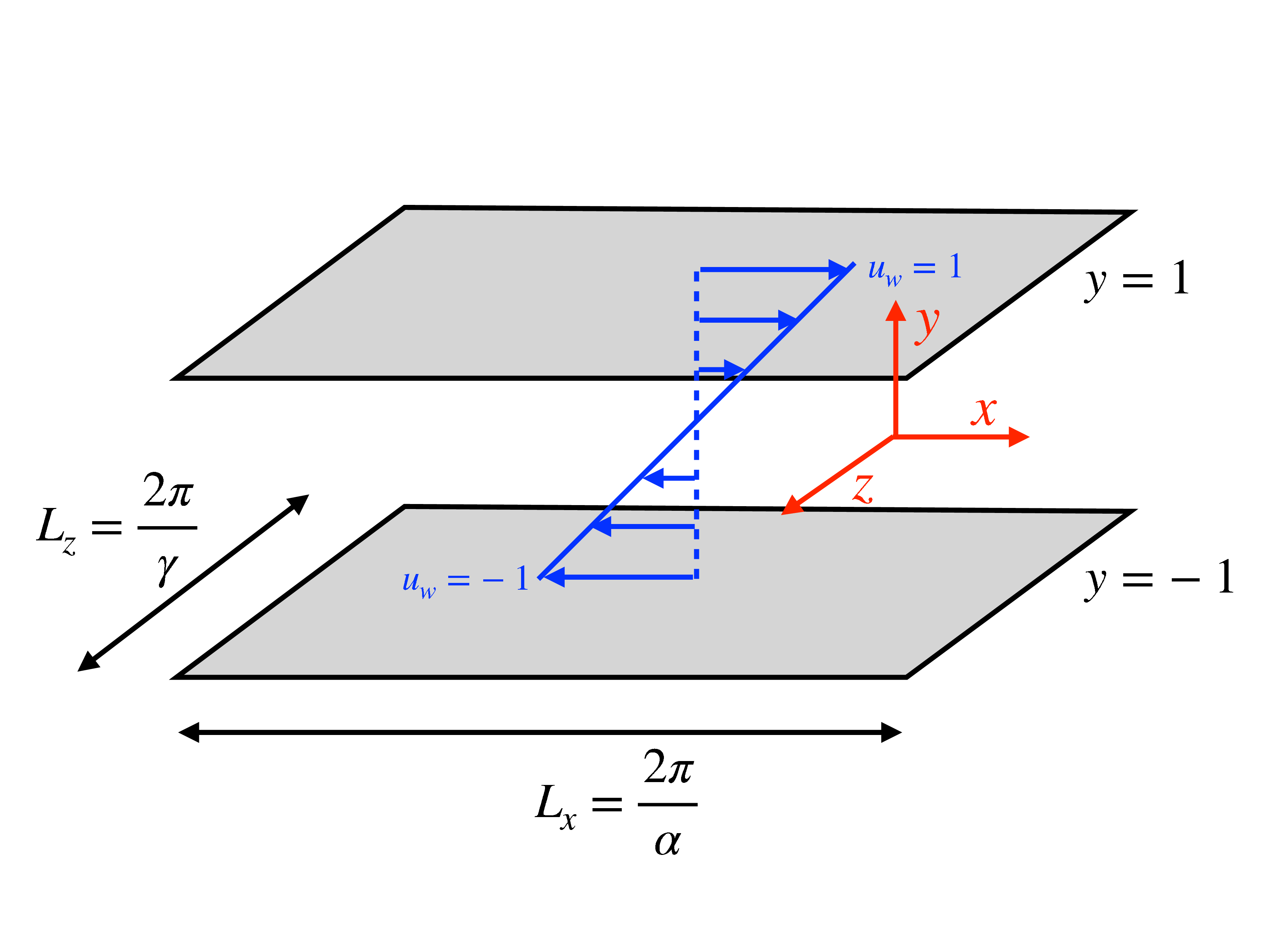}
\caption{Couette flow}
\end{subfigure}
\caption{Sketch of geometry, coordinate system (red) and laminar solutions (blue lines and arrows) for Waleffe and Couette flow.}
\label{fig:sketch}
\end{figure}

We consider here flows between two parallel walls, in a domain with lengths $(L_x,L_y,L_z)$ in streamwise, wall-normal and spanwise directions, respectively. Quantities are normalised by the half-channel height, and periodicity is assumed in streamwise and spanwise directions. This leads to a fundamental periodic box with lenghts $(L_x,L_y,L_z)  = (2\pi/\alpha, 2, 2\pi/\gamma)$, where $\alpha$ and $\gamma$ are fundamental wavenumbers in streamwise and spanwise directions. The flow is described using Cartesian coordinates $(x,y,z)$ denoting streamwise, wall-normal and spanwise directions, respectively, and $t$ representing time. The origin for Waleffe flow is taken at the lower wall, such that $y$ varies between 0 and 2, whereas for Couette flow the origin is more conveniently placed at the centre. The geometries and coordinate systems for Walefffe and Couette flow are sketched in figure \ref{fig:sketch}.

Waleffe flow~\citep{waleffe1997self,moehlis2004low,chantry2016turbulent} is a convenient shear flow for fundamental studies, as the application of free-slip conditions on the walls allows a straightforward use of Fourier modes to discretise all spatial directions. It is a shear flow forced by a body force in the streamwise direction 
\begin{equation}
\mathbf{f} = 
\begin{bmatrix}
f_x \\ f_y \\ f_z
\end{bmatrix}
= 
\begin{bmatrix}
-\frac{\sqrt{2}}{\mathrm{Re}} \cos(\beta y) \\ 0 \\ 0,
\end{bmatrix}
\end{equation}
where $\mathrm{Re}$ is the Reynolds number, and $\beta = \pi/2$ is a fundamental wall-normal wavenumber. Considering free-slip conditions on the walls at $y=0$ and $y=2$, this leads to a laminar solution $u(y) = \sqrt{2}\cos(\beta y)$, as in \citet{moehlis2004low}. This solution is linearly stable for all $\mathrm{Re}$ \citep{waleffe1997self}. The laminar solution is illustrated in the sketch of figure \ref{fig:sketch}.

\modif{In this work we will ultimately obtain a reduced-order model for Couette flow, as it allows comparisons with plenty of available numerical and experimental results. However, this will benefit from a first model for Waleffe flow, as free-slip boundary conditions allow a direct expansion of velocity components as Fourier modes. The strategy pursued here was to initially derive a Waleffe-flow model with desirable properties, and then transpose it to the Couette setup by an adaptation of the Waleffe basis to no-slip boundary conditions. This will be later explained in section \ref{sec:couettegalerkin}.}

We consider a velocity field $\mathbf{u} = \begin{bmatrix} u & v & w\end{bmatrix}^T$, where $u$, $v$ and $w$ denote respectively streamwise, wall-normal and spanwise velocity components. To obtain a reduced-order model for the Navier-Stokes system, we write the expansion $\mathbf{u}(x,y,z,t) = \sum_i {a_i(t) \mathbf{u}_i(x,y,z)}$, 
Following \citet{waleffe1997self} and \citet{moehlis2004low}, spatial modes $\mathbf{u}_i(x,y,z)$ are defined so as to satisfy periodic boundary conditions in $x$ and $z$, and free-slip conditions on the walls, $\partial u/\partial y = v = \partial w/\partial y = 0$ for $y=0$ and 2. In order to also satisfy the continuity equation, spatial modes are defined as
\begin{equation}
\mathbf{u}_i(x,y,z) = 
\begin{bmatrix}
u_i \\
v_i \\
w_i
\end{bmatrix}
=
\begin{bmatrix}
A_u(i) \sin(k_x(i) x + \phi_x(i)) \cos(k_y(i) y) \cos(k_z(i) z + \phi_z (i))  \\
A_v(i) \cos(k_x(i) x + \phi_x(i)) \sin(k_y(i) y) \cos(k_z(i) z + \phi_z (i))  \\
A_w(i) \cos(k_x(i) x + \phi_x(i)) \cos(k_y(i) y) \sin(k_z(i) z + \phi_z (i)) 
\end{bmatrix}
\label{eq:freeslipmodes}
\end{equation}
where the wavenumber of the mode is given by $\mathbf{k} = \begin{bmatrix}k_x &  k_y & k_z\end{bmatrix}^T$, with $k_x$, $k_y$ and $k_z$ as integer multiples of the fundamental wavenumbers $\alpha$, $\beta$ and $\gamma$. \modif{The amplitudes of the three velocity components are selected so as to ensure that modes form an orthonormal basis of divergence-free fields.} In what follows we avoid the notation with  $i$-dependence of amplitudes, wavenumbers and phases, and consider implicitly that we are dealing with mode $i$. $\phi_x$ and $\phi_z$ are phases in $x$ and $z$ directions, set as 0 or $\pi/2$ to ensure that modes are orthogonal. The amplitude vector $\mathbf{q} = \begin{bmatrix}A_u &  A_v & A_w\end{bmatrix}^T$ should be orthogonal to the wavenumber $\mathbf{k}$ to ensure incompressibility. This is ensured by considering a test wavenumber $\mathbf{k}_{test}=[0\; 0\; 1]^T$, and two amplitude vectors are obtained as $\mathbf{q}_1 = \mathbf{k} \times \mathbf{k}_{test}$ and  $\mathbf{q}_2 = \mathbf{k} \times \mathbf{q}_2$. These are by construction orthogonal to $\mathbf{k}$. If $\mathbf{k}$ is parallel to $\begin{bmatrix} 0 & 0 & 1\end{bmatrix}^T$ we use $\mathbf{k}_{test}=\begin{bmatrix}0 & 1 & 0\end{bmatrix}^T$ instead.

The procedure above was applied to wavenumbers going from $\begin{bmatrix} 0 & 0 & 0\end{bmatrix}^T$ to $\begin{bmatrix}3\alpha & 3\beta & 3\gamma\end{bmatrix}^T$, generating a large number of modes. Some combinations of amplitudes and phases lead to modes that vanish identically, and such modes are discarded. The remaining modes form an orthonormal basis with an inner product given by
\begin{equation}
\langle{\mathbf{f},\mathbf{g}} \rangle = \frac{1}{2L_x L_z}\int_0^{L_z}{\int_0^{2}{\int_0^{L_x}{\mathbf{f}(x,y,z) \cdot \mathbf{g}(x,y,z) \mathrm{d}x}\mathrm{d}y}\mathrm{d}z}
\end{equation}
Inserting 
\begin{equation}
\mathbf{u}(x,y,z,t) = \sum_j {a_j(t) \mathbf{u}_j(x,y,z)}
\label{eq:modaldecomposition}
\end{equation}
in the Navier-Stokes equation and taking an inner product with $\mathbf{u}_j$ leads to a system of ordinary differential equations of the form
\begin{equation}
\frac{\mathrm{d} a_i}{\mathrm{d} t}
=
F_i
+ \sum_j{L_{i,j}}a_j
+ \sum_j{\sum_k{Q_{i,j,k}}a_ja_k}
\end{equation}
with coefficients given by
\begin{equation}
F_i = \langle \mathbf{f},\mathbf{u_i} \rangle,
\end{equation}
\begin{equation}
L_{i,j} = \langle \nabla^2 \mathbf{u_j},\mathbf{u_i} \rangle,
\end{equation}
\begin{equation}
Q_{i,j,k} = -\langle (\mathbf{u_j} \cdot \nabla) \mathbf{u_k},\mathbf{u_i} \rangle.
\end{equation}

\andre{The procedure above is a Galerkin method to obtain a reduced-order model in the subspace spanned by the orthonormal modes. An important property of the method is that the error is orthogonal to the subspace. Thus, the non-linear interactions retained in a reduced basis are not an artefact of the projection, and are indeed present in the full Navier-Stokes system. However, the truncation of the description to a small number of modes restricts the number of possible mode interactions. \modif{For instance, if a wavenumber is not included in the basis all energy transfer mechanisms involving it are neglected.} This limits the accuracy of the resulting model, but, on the other hand, reduces the number of non-linear interactions between modes, which simplifies their study. In this work we search for a simple reduced-order model, which nonetheless leads to long turbulence lifetimes. The rationale to select a reduced basis is described in what follows.}

\subsection{Strategy for model reduction}
\label{sec:strategy}

\modif{The derivation of earlier reduced-order models by \citet{waleffe1997self} and \citet{moehlis2004low} was based on postulated linear and non-linear mechanisms for the dynamics of the flow, with modes selected in order to represent the formation of streaks by the lift-up mechanism, and a subsequent streak instability leading to a non-linear forcing of rolls. Here an \emph{a priori} assumption on dominant mechanisms is avoided, as the creation of a orthogonal large basis, described in the previous section, allows a straightforward truncation of the system to a small number of modes, whose dynamics may be studied by carrying out a handful of simulations. We thus select modes based on their role in maintaining longer turbulence lifetimes as determined by simulations of the Galerkin system. A recent study by \citet{lozano2020cause} reviews linear mechanisms postulated for wall-bounded turbulence, and results show that while all such mechanisms are plausible, only few of them are dominant in actual flow simulations; for instance, restriction of the system to streak transient growth, by an artificial removal of  streak instability mechanisms, leads to flows with statistics similar to results from full non-linear simulations. We thus avoid selection of modes based on a given mechanism to avoid doubts regarding its dominance or relevance in the flow; however, the resulting systems may be analysed \emph{a posteriori} to reveal the mechanisms at play in the reduced model.}

With wavenumbers up to $\begin{bmatrix} 3\alpha & 3\beta & 3\gamma \end{bmatrix}^T$ there are $4 \times 4 \times 4 \times 8 = 512$ possible modes if one considers the possible wavenumbers (4 in each direction, including the zero wavenumber), and the possible amplitudes (2) and phases (4). We have discarded from the set the vanishing modes, and also the two modes related to zero wavenumber in the three directions, to enforce zero mass flux at all times in both $x$ and $z$ directions. Once such modes are discarded from the set, the procedure of the last section led to a system of 342 ordinary differential equations for the evolution of the 342 mode amplitudes $a_i$. For such a large basis, we have employed a numerical quadrature based on spectral methods~\citep{weideman2000matlab,trefethen2000spectral} for a fast, but accurate derivation of Galerkin systems. We considered $L_x = 4\pi$ and $L_z = 2\pi$, one of the domain dimensions considered in the MFE model  \citep{moehlis2004low}. A time integration of this system is seen to lead to chaotic behaviour, with eventual relaminarisations, similar to the results of the MFE model. However, the typical lifetime of the transient chaos was observed to be at least an order of magnitude higher than the results reported by MFE.

We further constrained our model by reducing the number of modes in the basis. Such reduction had two constraints: the removal of a mode should not maintain the laminar solution as linearly stable, and neglecting a given mode should not drastically reduce the typical lifetime of chaotic behaviour.  The reduction was first carried out by considering only wavenumbers up to $\begin{bmatrix}\alpha & \beta & 2\gamma\end{bmatrix}^T$, reducing the basis to 44 modes. Following this, the basis was restricted to modes satisfying the $u,v,w(-x,1-y,-z) = -u,-v,-w(x,1+y,z)$ symmetry, as the modes in the MFE model. \modif{Imposing such symmetry fixes the streamwise and spanwise location fo structures, such that travelling waves and relative periodic orbits cannot be obtained in the reduced-order model, simplifying the study of the dynamics, as in \citet{kreilos2014increasing}. Such symmetry}  led to a further reduction to 30 modes. Both reductions were seen to have low impact on the lifetime of chaotic periods. 

This last basis was sufficiently reduced to allow a final, manual reduction of the system by neglecting modes based on trial and error.  This led to a system with 12 modes, reported in table \ref{tab:modes}. Among these basis functions, eight modes correspond closely to the basis used by \citet{waleffe1997self}, and  reappear in modified form in the MFE model. \modif{Mode 4 is a streak, with streamwise constant fluctuations of the streamwise velocity $u$, with spanwise wavelength equal to the domain size $L_z$. Mode 5 represents a roll, or streamwise vortex, with streamwise constant fluctuations of $v$ and $w$, also with spanwise wavelength of $L_z$.} Modes 11 and 12 are streaks and rolls, but with spanwise wavenumber equal to $2\gamma$, or, equivalently, spanwise wavelength of $L_z/2$. In the Waleffe and MFE models two oblique modes, with wavenumber $\mathbf{k}=\begin{bmatrix} \alpha & \beta & \gamma \end{bmatrix}$, were included in the basis. Here three modes were retained, and linear combinations of these three modes lead to the functions in the earlier models; hence, only one additional degree of freedom is introduced for this wavenumber. \modif{Such additional mode is marked in table $\ref{tab:modes}$ as related to mode 9. Modes 8 and 9 only differ in their phases, and mode 9 may be considered as an addition of the present model as it may be neglected from the model without leading to a linear instability of the laminar solution}. The present basis comprises two wall-normal vortex modes 6 and 7, which identical except for phase shifts in $x$ and $z$. The earlier works only included a single mode representing $y$-vortices.  On the other hand, the mean-flow distortion mode with vertical wavenumber $3\beta$ in the MFE model \modif{(mode 9 in the notation of that work)} was not retained in the present reduction procedure, \modif{as it did not lead to significant changes to chaotic lifetimes. Another difference with respect to the MFE model is that modes with a $\cos^2(\pi y/2)$ dependence were used in that work, involving thus more than one wavenumber per mode. Therefore, the present ROM cannot be reduced to the MFE model by neglecting modes in the basis, but it is expected that the removal of modes 7, 9, 11 and 12, plus the addition of a new mean-flow distortion mode, would lead to similar behaviour to the observations in MFE.}

\begin{table}
  \begin{center}
  \begin{tabular}{|c|c|c|c|c|c|c|c|c|c|}
  \hline
	Mode & $k_x/\alpha$ & $k_y/\beta$ & $k_z/\gamma$ & $A_u$ & $A_v$ & $A_w$ & $\frac{\phi_x}{(\pi/2)}$ & $\frac{\phi_z}{(\pi/2)}$ & Structure \\ \hline
	1 (M) & 0 & 1 & 0 & $-\sqrt{2}$ & 0 & 0 & 1 & 0 & Mean flow  \\ 
	2 (A) & 1 & 0 & 0 & 0 & 0 & $-\sqrt{2}$ & 1 & 1 & Even spanwise flows \\
	3 (C) & 1 & 1 & 0 & 0 & 0 & $-2$ & 0 & 1 &  {Odd spanwise flows} \\
	4 (U) & 0 & 0 & 1 & $\sqrt{2}$ & 0 & 0 & 1 & 1 & Streaks \\
	5 (V) & 0 & 1 & 1 & 0 & $2\frac{\gamma}{k_{\beta\gamma}}$ & $-2\frac{\beta}{k_{\beta\gamma}}$ & 0 & 1 & Rolls \\
	6 (B) & 1 & 0 & 1 & $2\frac{\gamma}{k_{\alpha\gamma}}$ & 0 & $-2\frac{\alpha}{k_{\alpha\gamma}}$ & 0 & 0 & $y$-vortices 1 \\
	\textbf{7} & 1 & 0 & 1 & $2\frac{\gamma}{k_{\alpha\gamma}}$ & 0 & $-2\frac{\alpha}{k_{\alpha\gamma}}$ & 1 & 1 & \textbf{$y$-vortices 2} \\
	{8} & 1 & 1 & 1 & $-2 \sqrt{2} \frac{\beta}{k_{\alpha\beta}}$ & $2 \sqrt{2} \frac{\alpha}{k_{\alpha\beta}}$ & 0 & 1 & 0 & {Oblique wave 1} \\
	\textbf{9} & 1 & 1 & 1 & $-2 \sqrt{2} \frac{\beta}{k_{\alpha\beta}}$ & $2 \sqrt{2} \frac{\alpha}{k_{\alpha\beta}}$ & 0 & 0 & 1 & \textbf{Oblique wave 2} \\
	10 & 1 & 1 & 1 & $2\alpha\gamma/N$ & $2\beta \gamma/N$ & $-2k_{\alpha,\beta}^2/N$ & 1 & 0 & Oblique wave 3 \\
	\textbf{11} & 0 & 0 & 2 & $\sqrt{2}$ & 0 & 0 & 1 & 1 & \textbf{$L_z/2$ streaks} \\	
	\textbf{12} & 0 & 1 & 2 & 0 & $4\frac{\gamma}{k_{\beta,2\gamma}}$ & $-2{\beta}{k_{\beta,2\gamma}}$ & 0 & 1 & \textbf{$L_z/2$ rolls}  \\	\hline
  \end{tabular}
  \caption{Modes in the Galerkin system for Waleffe flow. The normalisation constant for mode 10 is given by $N=\sqrt{\left(\alpha ^2+\beta ^2\right)\,\left(\alpha ^2+\beta ^2+\gamma ^2\right)/2}$. Auxiliary wavenumbers $k_{\alpha,\beta}$, $k_{\alpha,\gamma}$ and so on are defined in the text. Modes absent from the Waleffe and MFE models are highlighted in bold font. Corresponding modes in the Waleffe model are marked with letters (M, U, V, A, B, C).}
\label{tab:modes}
  \end{center}
\end{table}

Once the basis was reduced to 12 modes, the model coefficients could be obtained directly by integration of the basis functions and their derivatives, avoiding the numerical quadratures used in the initial steps. The system of differential equations of the present model of Waleffe flow is given by

\begin{subequations}
\begin{eqnarray}
\frac{\mathrm{d} a_1}{\mathrm{d} t}=
\frac{\beta \,\left(\beta -a_{1}\,\beta \right)}{\mathrm{Re}}+\frac{a_{4}\,a_{5}\,\beta \,\gamma }{k_{\beta,\gamma}}+\frac{2\,a_{11}\,a_{12}\,\beta \,\gamma }{k_{\beta,2\gamma}}-\frac{a_{6}\,a_{10}\,\beta ^2\,\gamma ^2}{k_{\alpha,\beta}\,k_{\alpha,\gamma}\,k_{\alpha,\beta,\gamma}}\nonumber \\
-\frac{a_{6}\,a_{8}\,\alpha \,\beta \,\gamma }{k_{\alpha,\beta}\,k_{\alpha,\gamma}}+\frac{a_{7}\,a_{9}\,\alpha \,\beta \,\gamma }{k_{\alpha,\beta}\,k_{\alpha,\gamma}}
\end{eqnarray}
\begin{eqnarray}
\frac{\mathrm{d} a_2}{\mathrm{d} t}=
-\frac{a_{2}\,\alpha ^2}{\mathrm{Re}} + a_{1}\,a_{3}\,\alpha +\frac{a_{4}\,a_{6}\,\alpha ^2}{k_{\alpha,\gamma}}+\frac{a_{5}\,a_{8}\,\alpha \,\beta ^2}{k_{\alpha,\beta}\,k_{\beta,\gamma}}-\frac{a_{5}\,a_{10}\,\alpha ^2\,\beta \,\gamma }{k_{\alpha,\beta}\,k_{\beta,\gamma}\,k_{\alpha,\beta,\gamma}}
\end{eqnarray}
\begin{eqnarray}
\frac{\mathrm{d} a_3}{\mathrm{d} t}=
-\frac{a_{3}\,\left(\alpha ^2+\beta ^2\right)}{\mathrm{Re}}-a_{1}\,a_{2}\,\alpha -\frac{a_{4}\,a_{10}\,\alpha \,k_{\alpha,\beta}}{k_{\alpha,\beta,\gamma}}
\end{eqnarray}
\begin{eqnarray}
\frac{\mathrm{d} a_4}{\mathrm{d} t}=
-\frac{a_{4}\,\gamma ^2}{\mathrm{Re}}+\frac{a_{3}\,a_{10}\,\alpha \,\gamma ^2}{k_{\alpha,\beta}\,k_{\alpha,\beta,\gamma}}-\frac{a_{2}\,a_{6}\,\gamma ^2}{k_{\alpha,\gamma}}-\frac{a_{3}\,a_{8}\,\beta \,\gamma }{k_{\alpha,\beta}}-\frac{a_{1}\,a_{5}\,\beta \,\gamma }{k_{\beta,\gamma}}
\end{eqnarray}
\begin{eqnarray}
\frac{\mathrm{d} a_5}{\mathrm{d} t}=
-\frac{a_{5}\,\left(\beta ^2+\gamma ^2\right)}{\mathrm{Re}}+\frac{2\,a_{3}\,a_{6}\,\alpha \,\beta \,\gamma }{k_{\alpha,\gamma}\,k_{\beta,\gamma}}-\frac{a_{2}\,a_{8}\,\alpha \,\left(\beta ^2-\gamma ^2\right)}{k_{\alpha,\beta}\,k_{\beta,\gamma}} \nonumber \\
+\frac{a_{2}\,a_{10}\,\beta \,\gamma \,\left(2\,\alpha ^2+\beta ^2+\gamma ^2\right)}{k_{\alpha,\beta}\,k_{\beta,\gamma}\,k_{\alpha,\beta,\gamma}}
\end{eqnarray}
\begin{eqnarray}
\frac{\mathrm{d} a_6}{\mathrm{d} t}=
-\frac{a_{6}\,\left(\alpha ^2+\gamma ^2\right)}{\mathrm{Re}} + \frac{2\,a_{1}\,a_{8}\,\alpha \,\beta \,\gamma }{k_{\alpha,\beta}\,k_{\alpha,\gamma}}-\frac{a_{2}\,a_{4}\,\left(\alpha ^2-\gamma ^2\right)}{k_{\alpha,\gamma}} \nonumber \\
-\frac{\sqrt{2}\,a_{7}\,a_{11}\,\alpha \,\left(\alpha ^2-3\,\gamma ^2\right)}{2\,\left(\alpha ^2+\gamma ^2\right)}
-\frac{a_{1}\,a_{10}\,\left(\alpha ^4+\alpha ^2\,\beta ^2+\alpha ^2\,\gamma ^2-\beta ^2\,\gamma ^2\right)}{k_{\alpha,\beta}\,k_{\alpha,\gamma}\,k_{\alpha,\beta,\gamma}}\nonumber \\
-\frac{2\,a_{3}\,a_{5}\,\alpha \,\beta \,\gamma }{k_{\alpha,\gamma}\,k_{\beta,\gamma}}+\frac{\sqrt{2}\,a_{9}\,a_{12}\,\beta ^2\,\left(\alpha ^2-\gamma ^2\right)}{2\,k_{\beta,2\gamma}\,k_{\alpha,\beta}\,k_{\alpha,\gamma}}
\end{eqnarray}
\begin{eqnarray}
\frac{\mathrm{d} a_7}{\mathrm{d} t}=
-\frac{a_{7}\,\left(\alpha ^2+\gamma ^2\right)}{\mathrm{Re}}+\frac{\sqrt{2}\,a_{6}\,a_{11}\,\alpha \,\left(\alpha ^2-3\,\gamma ^2\right)}{2\,\left(\alpha ^2+\gamma ^2\right)}-\frac{2\,a_{1}\,a_{9}\,\alpha \,\beta \,\gamma }{k_{\alpha,\beta}\,k_{\alpha,\gamma}} \nonumber \\
+\frac{\sqrt{2}\,a_{8}\,a_{12}\,\beta ^2\,\left(\alpha ^2-\gamma ^2\right)}{2\,k_{\beta,2\gamma}\,k_{\alpha,\beta}\,k_{\alpha,\gamma}}-\frac{\sqrt{2}\,a_{10}\,a_{12}\,\alpha \,\beta \,\gamma \,\left(3\,\alpha ^2+2\,\beta ^2-\gamma ^2\right)}{2\,k_{\beta,2\gamma}\,k_{\alpha,\beta}\,k_{\alpha,\gamma}\,k_{\alpha,\beta,\gamma}}
\end{eqnarray}
\begin{eqnarray}
\frac{\mathrm{d} a_8}{\mathrm{d} t}=
-\frac{a_{8}\,\left(\alpha ^2+\beta ^2+\gamma ^2\right)}{\mathrm{Re}}+\frac{a_{3}\,a_{4}\,\beta \,\gamma }{k_{\alpha,\beta}}-\frac{\sqrt{2}\,a_{9}\,a_{11}\,\alpha }{2}-\frac{a_{2}\,a_{5}\,\alpha \,\gamma ^2}{k_{\alpha,\beta}\,k_{\beta,\gamma}}
\nonumber \\
-\frac{a_{1}\,a_{6}\,\alpha \,\beta \,\gamma }{k_{\alpha,\beta}\,k_{\alpha,\gamma}}-\frac{\sqrt{2}\,a_{7}\,a_{12}\,\gamma ^2\,\left(4\,\alpha ^2-\beta ^2\right)}{2\,k_{\beta,2\gamma}\,k_{\alpha,\beta}\,k_{\alpha,\gamma}}
\end{eqnarray}
\begin{eqnarray}
\frac{\mathrm{d} a_9}{\mathrm{d} t}=
-\frac{a_{9}\,\left(\alpha ^2+\beta ^2+\gamma ^2\right)}{\mathrm{Re}}+\frac{\sqrt{2}\,a_{8}\,a_{11}\,\alpha }{2}-\frac{\sqrt{2}\,a_{10}\,a_{11}\,\beta \,\gamma }{k_{\alpha,\beta,\gamma}}\nonumber \\
+\frac{a_{1}\,a_{7}\,\alpha \,\beta \,\gamma }{k_{\alpha,\beta}\,k_{\alpha,\gamma}}-\frac{\sqrt{2}\,a_{6}\,a_{12}\,\gamma ^2\,\left(4\,\alpha ^2-\beta ^2\right)}{2\,k_{\beta,2\gamma}\,k_{\alpha,\beta}\,k_{\alpha,\gamma}}
\end{eqnarray}
\begin{eqnarray}
\frac{\mathrm{d} a_{10}}{\mathrm{d} t}=
-\frac{a_{10}\,\left(\alpha ^2+\beta ^2+\gamma ^2\right)}{\mathrm{Re}}+\frac{a_{1}\,a_{6}\,\alpha ^2\,k_{\alpha,\beta,\gamma}}{k_{\alpha,\beta}\,k_{\alpha,\gamma}}+\frac{a_{3}\,a_{4}\,\alpha \,\left(\alpha ^2+\beta ^2-\gamma ^2\right)}{k_{\alpha,\beta}\,k_{\alpha,\beta,\gamma}}\nonumber \\
-\frac{a_{2}\,a_{5}\,\beta \,\gamma \,k_{\alpha,\beta,\gamma}}{k_{\alpha,\beta}\,k_{\beta,\gamma}}-\frac{\sqrt{2}\,a_{7}\,a_{12}\,\alpha \,\beta \,\gamma \,\left(\alpha ^2+\beta ^2+5\,\gamma ^2\right)}{2\,k_{\beta,2\gamma}\,k_{\alpha,\beta}\,k_{\alpha,\gamma}\,k_{\alpha,\beta,\gamma}}
\end{eqnarray}
\begin{eqnarray}
\frac{\mathrm{d} a_{11}}{\mathrm{d} t}=
-\frac{4\,a_{11}\,\gamma ^2}{\mathrm{Re}}-\gamma \,\left(\frac{2\,a_{1}\,a_{12}\,\beta }{k_{\beta,2\gamma}}-\frac{\sqrt{2}\,a_{9}\,a_{10}\,\beta }{k_{\alpha,\beta,\gamma}}\right)
\end{eqnarray}
\begin{eqnarray}
\frac{\mathrm{d} a_{12}}{\mathrm{d} t}=
-\frac{a_{12}\,\left(\beta ^2+4\,\gamma ^2\right)}{\mathrm{Re}}
+\frac{\sqrt{2}\,a_{7}\,a_{10}\,\alpha \,\beta \,\gamma \,\left(4\,\alpha ^2+3\,\beta ^2+4\,\gamma ^2\right)}{2\,k_{\beta,2\gamma}\,k_{\alpha,\beta}\,k_{\alpha,\gamma}\,k_{\alpha,\beta,\gamma}}\nonumber \\
-\frac{\sqrt{2}\,a_{6}\,a_{9}\,\alpha ^2\,\left(\beta ^2-4\,\gamma ^2\right)}{2\,k_{\beta,2\gamma}\,k_{\alpha,\beta}\,k_{\alpha,\gamma}}
-\frac{\sqrt{2}\,a_{7}\,a_{8}\,\alpha ^2\,\left(\beta ^2-4\,\gamma ^2\right)}{2\,k_{\beta,2\gamma}\,k_{\alpha,\beta}\,k_{\alpha,\gamma}},
\end{eqnarray}
\label{eq:waleffemodel}
\end{subequations}
with auxiliary wavenumbers $k_{\alpha,\beta} = \sqrt{\alpha^2 + \beta^2}$, $k_{\beta,\gamma} = \sqrt{\beta^2 + \gamma^2}$, $k_{\alpha,\gamma} = \sqrt{\alpha^2 + \gamma^2}$, $k_{\alpha,\beta,\gamma} = \sqrt{\alpha^2 + \beta^2+\gamma^2}$ and $k_{\beta,2\gamma} = \sqrt{\beta^2 + 4 \gamma^2}$. A numerical solution to the system is possible by starting with an initial condition to the mode coefficients $a_i$ and advancing with the Runge-Kutta method, for instance. Here a standard Runge-Kutta method of 4th/5th order was applied. From the time series of the mode coefficients, the velocity field may be recovered using eq. (\ref{eq:modaldecomposition}).

In the model equations, the only forced coefficient is $a_1$, all coefficients are damped by the viscous term (first term in the right-hand side) and quadratic terms conserve energy, only redistributing it among the modes. The laminar solution is $a_1=1$, $a_2=a_3=...=a_{12}=0$, corresponding to $u_0=-\sqrt{2}\cos(\beta y)$, as in the MFE model. Notice that here, instead of a sine, the laminar solution is a cosine with minus sign due to the position of walls at $y=0$ and $y=2$ (as shown in figure \ref{fig:sketch}\textit{a}).

Inspection of the model shows some terms that may be directly related to the Waleffe (here taken in its 8-mode version) and MFE models. In the equation for the fundamental streak amplitude $a_4$, we observe a ``lift-up'' term proportional to $a_1 a_5$. Non-zero rolls $a_5$ may lead to algebraic growth of the streak $a_4$ in the presence of mean shear $a_1$. A similar term appears in the equation of the $L_z/2$ streak $a_{11}$, with a lift-up term proportional to $a_1 a_{12}$ related to the $L_z/2$ roll $a_{12}$. Other non-linear interactions are not as evident from model inspection, but comparison with the Waleffe model shows that the $a_2a_6$ term is one of the terms describing streak instability (the $AB$ term in the equation for $U$); other terms differ due to the choice of oblique waves in the present model. The fundamental rolls $a_5$ are excited by the non-linear interaction $a_3a_6$ that matches the $BC$ term in the Waleffe model, and thus mode 6, which comprises wall-normal vortices, is involved in both streak instability and regeneration of rolls. Mode 6 also appears in the equation for the $L_z/2$ rolls $a_{12}$; notice that the other mode describing wall-normal vortices, mode 7, excites the $L_z/2$ rolls.

\modif{The appearance of streaks and rolls at wavelengths of $L_z$ and $L_z/2$ may be related to observations in some recent works. The restricted non-linear system by \citet{farrell2012dynamics} and \citet{thomas2015minimal} shows that it is possible to greatly truncate non-linear interactions with higher streamwise wavenumber and maintain statistics similar to the full Navier-Stokes system, provided all spanwise wavenumbers are considered. \citet{lozano2020cause} have also performed various truncations of the system, and the key process in wall-bounded turbulence was shown to be related to transient growth of disturbances growing on a streaky base flow. Such transient growth is mostly associated with the Orr mechanism, related to the spanwise shear introduced by the streaks. The inclusion of $L_z/2$ streaks in the present ROM enhances the possibilities for such transient growth, as mode 11 leads to higher spanwise shear.}

Thus, the model has features that could, in principle, lead to cycles similar to the one studied by \citet{hamilton1995regeneration}, for a spanwise wavelength of $L_z$, but also of $L_z/2$. However, a simple inspection of the model does not show how these \modif{wavelengths} may be related. This will be investigated in further detail when analysing the results of the model.

\subsection{Adaptation of the model to Couette flow}
\label{sec:couettegalerkin}

To adapt the model of eq. (\ref{eq:waleffemodel}) to plane Couette flow between two horizontal walls with opposite velocities, some changes are necessary. The first is the consideration of a decomposition into laminar solution and fluctuations,
\begin{equation}
\mathbf{u}(\mathbf{x},t) = \mathbf{u}_0(\mathbf{x}) + \mathbf{u}'(\mathbf{x},t)
\end{equation}
where $\mathbf{u}_0(\mathbf{x})=\begin{bmatrix}u_0(y) & 0 & 0\end{bmatrix}^T=\begin{bmatrix}y & 0 & 0\end{bmatrix}$ is the laminar solution satisfying boundary conditions $u(\pm 1) = \pm 1$ at walls; the wall velocity is used here as the reference velocity. Notice that for Couette flow the walls are more conveniently placed at $y=\pm 1$, as sketched in figure \ref{fig:sketch}(b). 

We consider the fluctuations around the laminar solution to be written as $\mathbf{u}'(\mathbf{x},t) = \sum_i a_i(t) \mathbf{u}_i(\mathbf{x})$, where the modes $\mathbf{u}_i(\mathbf{x})$ satisfy non-slip conditions on the walls. The free-slip modes in eq. (\ref{eq:freeslipmodes}) are no longer appropriate, and Fourier modes in $y$ are replaced by polynomials following \citet{lagha2007modeling}. This leads to
\begin{equation}
\mathbf{u}_i(x,y,z) = 
\begin{bmatrix}
u_i \\
v_i \\
w_i
\end{bmatrix}
=
\begin{bmatrix}
A_u(i) \sin(k_x(i) x + \phi_x(i)) (1-y^2) \cos(k_z(i) z + \phi_z (i))  \\
0  \\
A_w(i) \cos(k_x(i) x + \phi_x(i)) (1-y^2) \sin(k_z(i) z + \phi_z (i)) 
\end{bmatrix}
\label{eq:evenmodes}
\end{equation}
for modes that are even around $y=0$ for $u$ and $w$, which correspond to $k_y=0$; and 
\begin{equation}
\mathbf{u}_i(x,y,z) = 
\begin{bmatrix}
u_i \\
v_i \\
w_i
\end{bmatrix}
=
\begin{bmatrix}
A_u(i) \sin(k_x(i) x + \phi_x(i)) \frac{4}{\beta}y(1-y^2) \cos(k_z(i) z + \phi_z (i))  \\
A_v(i) \cos(k_x(i) x + \phi_x(i)) (1-y^2)^2 \cos(k_z(i) z + \phi_z (i))   \\
A_w(i) \cos(k_x(i) x + \phi_x(i)) \frac{4}{\beta}y(1-y^2) \sin(k_z(i) z + \phi_z (i)) 
\end{bmatrix}
\label{eq:oddmodes}
\end{equation}
for modes that are odd around $y=0$ for $u$ and $w$, corresponding to $k_y=\beta$. The polynomials in $y$ satisfy the non-slip conditions requiring $u=v=w=0$ on the walls; notice that the first $y$ derivative of $v$ also vanishes, as imposed by the contiuity equation. If we consider $\beta=\sqrt{3}$ the same modes of table~\ref{tab:modes} can be used as an orthogonal basis, which may subsequently be normalised in straightforward manner. The orthonormal basis is shown in eq. (\ref{eq:couettemodes}) in the Appendix. Such direct use of the modes obtained in the system truncation for Waleffe flow implicitly considers the similarity between these two flows in the central region of the channel, as observed by~\citet{chantry2016turbulent}.

A Galerkin projection is applied for the Navier-Stokes system applied to $\mathbf{u'}$, which leads to a modified linear operator,
\begin{equation}
L_{i,j} = \langle \nabla^2 \mathbf{u_j},\mathbf{u_i} \rangle - Re\langle \left[(\mathbf{u_j} \cdot \nabla) \mathbf{u_0} + (\mathbf{u_0} \cdot \nabla) \mathbf{u_j} \right],\mathbf{u_i} \rangle,
\end{equation}
no change in the quadratic term and $F_i=0$. Such Galerkin projection of velocity fluctuations was verified by application to Waleffe flow, leading to the same statistics of the Galerkin system of the total velocity. 

A system of twelve ordinary differential equations for Couette flow is given for convenience in eq. \ref{eq:couettemodel} in Appendix \ref{sec:equations}, as the equations become lengthy. The system is structurally similar to the model for Waleffe flow, but here the linear term includes as well couplings between the modes and the laminar solution. \modif{Notice that for such linear terms with coupling to the laminar solution there is a corresponding quadratic term showing coupling to mode 1, which for Couette flow represents mean-flow distortion; thus, such linear and quadratic terms may be thought in combination as related to a mean-flow effect.} The laminar solution for Couette flow is recovered with zero fluctuations, implying $a_1=a_2=...=a_{12}=0$.

\modif{It is worth emphasising that modes equivalent to the ones in the Waleffe ROM were used for Couette flow, with the insight that the two flows display similarities~\cite{chantry2016turbulent}. Thus, the reduced basis obtained initially for Waleffe flow was directly adapted for the Couette configuration, ensuring an equivalence between the two ROMs.}

\section{Model results}
\label{sec:results}

\subsection{Waleffe flow}

We first explore the reduced-order model of Waleffe flow in eq. (\ref{eq:waleffemodel}). Throughout this work we consider $\alpha=0.5$ and $\gamma=1$, which leads to a numerical box with $L_x=4\pi$ and $L_z=2\pi$. This is one of the domains considered by \citet{moehlis2004low}. Other choices of computational domain did not lead to major changes in the results, as exemplified for Couette flow in Appendix \ref{sec:boxsize}.  Figure \ref{fig:waleffetimeseries} shows time series of two sample runs of the model starting from different random initial conditions, considering $Re=200$. Similar to the MFE model, the mode-1 amplitude $a_1$ is seen to approach the laminar value $a_1=1$; in the simulation of figure \ref{fig:waleffetimeseries}(a) the model relaminarises, whereas in figure \ref{fig:waleffetimeseries}(b) the chaotic behaviour persists up to $t=10^5$. Similar behaviour is observed for other Reynolds numbers, but with different lifetimes of chaotic behaviour. As in the MFE model, the present ROM does not present sustained turbulence. This is likely due to the small computational domain, as discussed in the Introduction, but also due to the severe truncation of the system. However, the observed lifetimes are significantly higher than what is observed for the MFE model. For $Re=200$ \citet{moehlis2004low} report a median lifetime of approximately 1000, which is much lower than the observations of the present system. This will be more accurately quantified in what follows.

\begin{figure}
\begin{subfigure}{0.49\textwidth}
\includegraphics[width=1.0\textwidth]{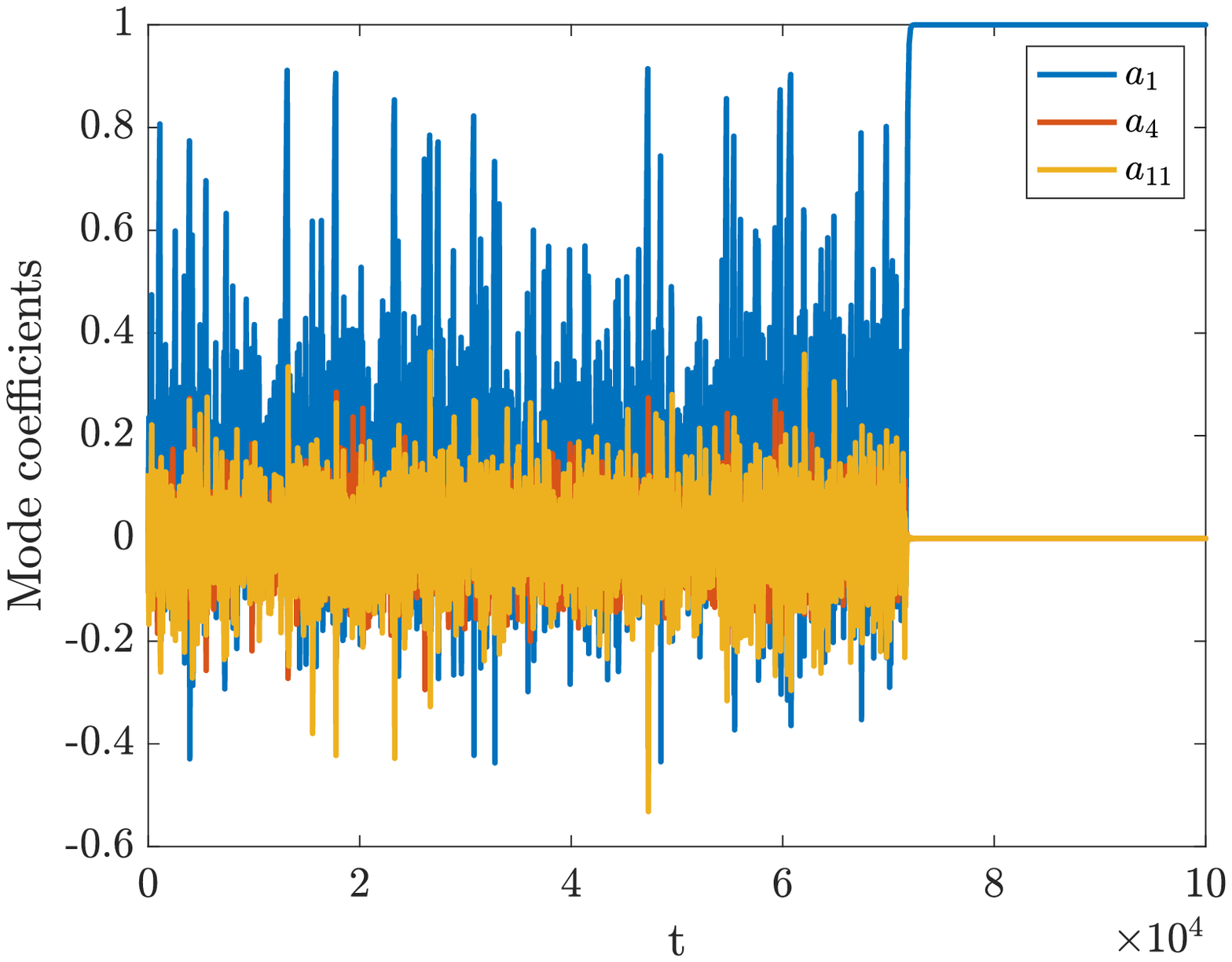}
\caption{Sample run with relaminarisation}
\end{subfigure}
\begin{subfigure}{0.49\textwidth}
\includegraphics[width=1.0\textwidth]{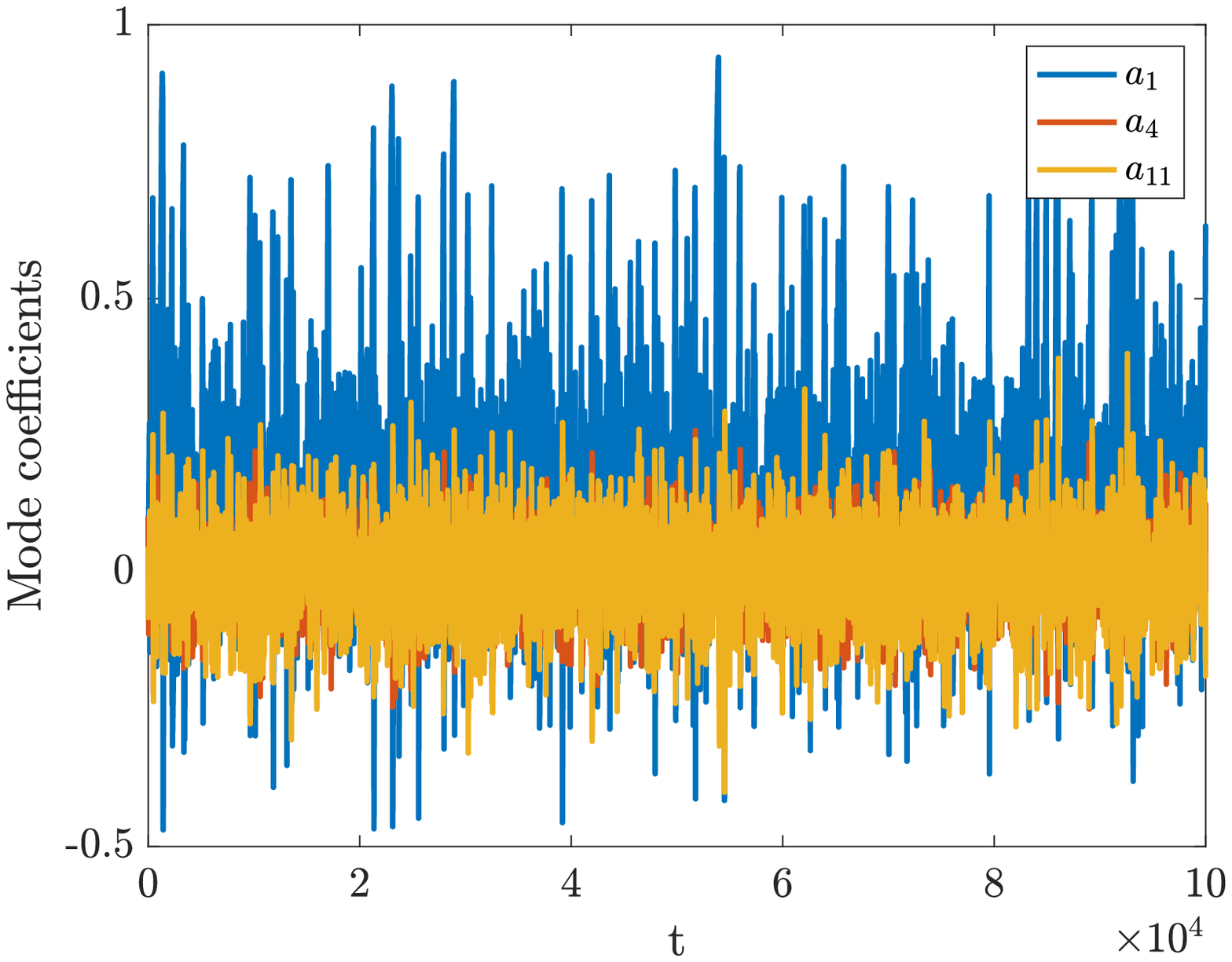}
\caption{Sample run without relaminarisation}
\end{subfigure}
\caption{Sample model results for $\mathrm{Re}=200$. The temporal coefficients of selected modes, $a_1$ (mean flow), $a_4$ (streaks) and $a_{11}$ ($L_z/2$ streaks) are shown.}
\label{fig:waleffetimeseries}
\end{figure}

\begin{figure}
\begin{subfigure}{0.49\textwidth}
\includegraphics[width=1.0\textwidth]{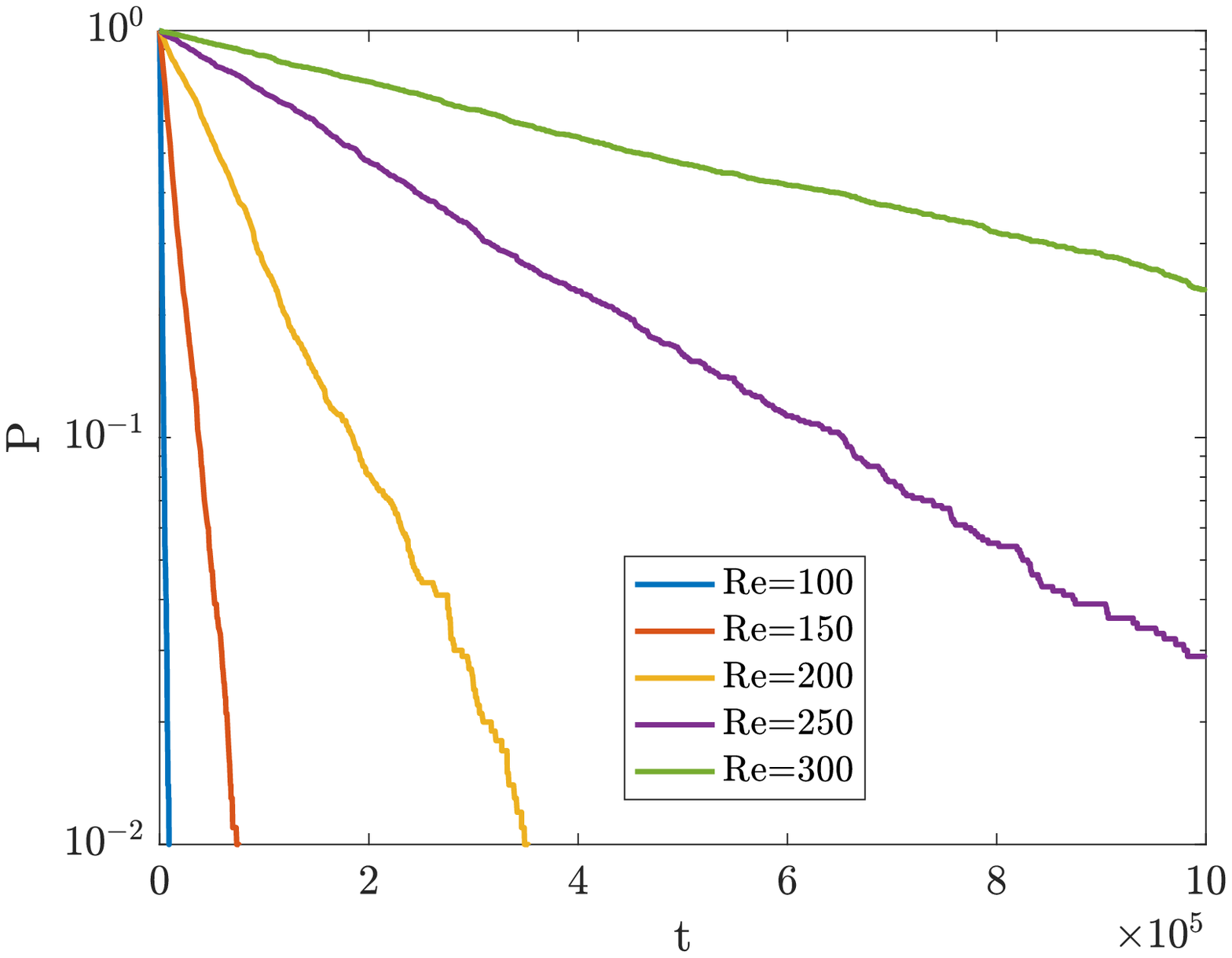}
\caption{Probability of chaotic behaviour until time $t$, $P(t)$}
\end{subfigure}
\begin{subfigure}{0.49\textwidth}
\includegraphics[width=1.0\textwidth]{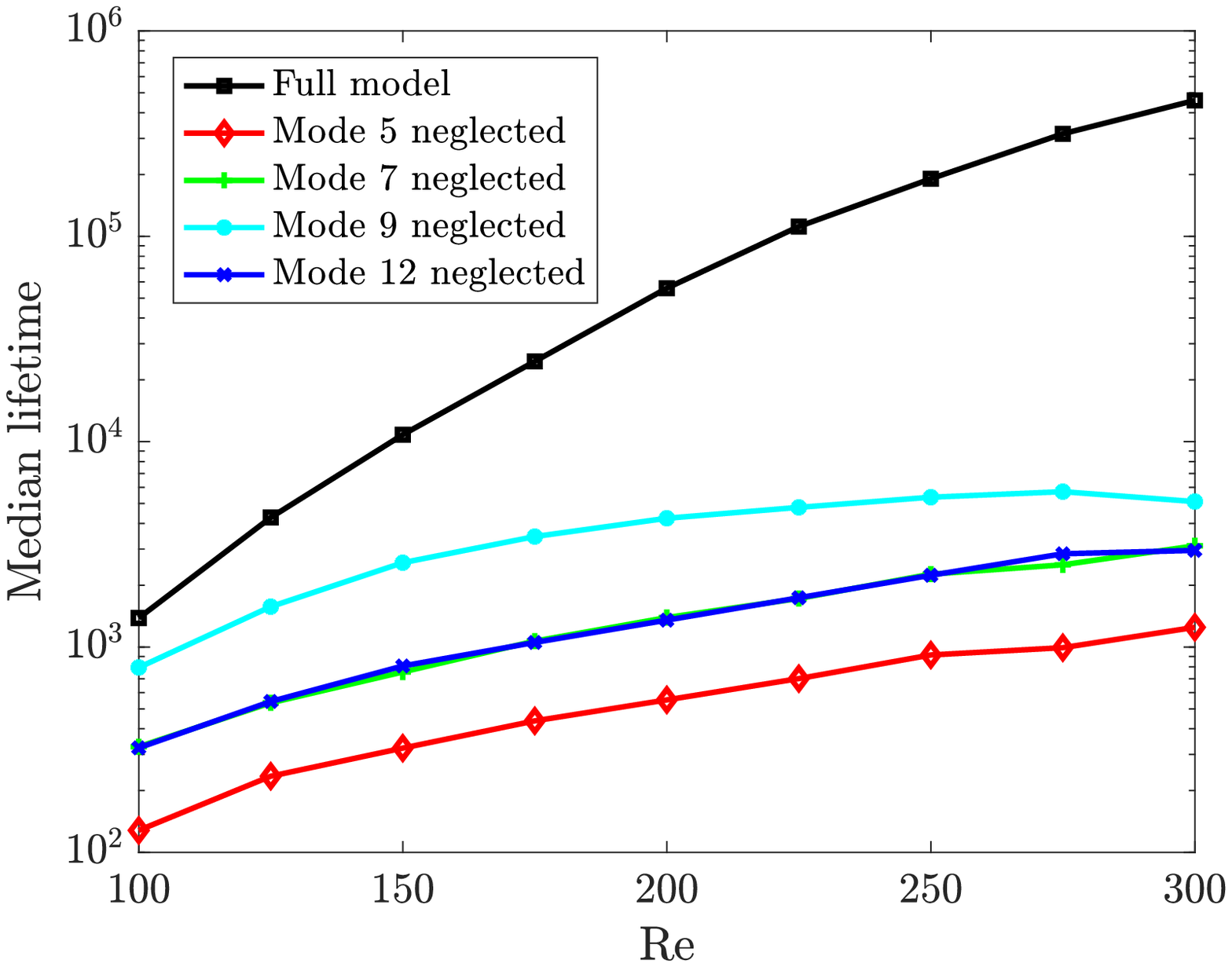}
\caption{Median lifetime as a function of Reynolds number}
\end{subfigure}
\caption{Waleffe-flow model lifetime statistics, taken from 1000 simulations with random initial conditions with norm equal to 0.3.}
\label{fig:waleffelifetimes}
\end{figure}

Lifetimes of chaotic behaviour may be systematically studied by running a large number of simulations with random initial conditions in order to determine the probability $P(t)$ of chaotic behaviour after time $t$, as in \citet{bottin1998statistical} and \citet{moehlis2004low}. 1000 simulations were ran for each Reynolds number with random initial conditions satisfying $\sum_i{a'_i}^2=0.09$, where $a'_i$ denotes a perturbation from the laminar solution ($a'_1=a_1-1, a'_i=a_i$ for $i\ge 2$). The system was considered to achieve the laminar state at time $t$ if $\sum_i{(a'_i)^2}<0.01$. The probability $P(t)$ is shown in figure \ref{fig:waleffelifetimes}(a) for Reynolds numbers between 100 and 300.  For a given Reynolds number, $P(t)$ decays exponentially with increasing $t$, and higher Reynolds numbers have slower decay rates. Such exponential decay of $P(t)$ is consistent with findings for the canonical wall-bounded flows \citep{bottin1998statistical,willis2007critical,kreilos2014increasing}, as well as for the MFE model, and suggests a memoryless process. This can be further characterised by the median lifetime as a function of $Re$, shown in figure  \ref{fig:waleffelifetimes}(b). As the Reynolds number is increased from 100 to 300, the median lifetime increases almost three orders of magnitude, and gets close to $10^6$ for $Re=300$.

Figure \ref{fig:waleffelifetimes}(b) also shows results of turbulence lifetimes for the model when modes 5 (fundamental rolls) or 12 ($L_z/2$ rolls) are neglected. The impact is substantial, and neglecting either one of the modes leads to a reduction of more than an order of magnitude in turbulence lifetimes. Such lower lifetimes have the same order of magnitude of the values reported by \citet{moehlis2004low}, which suggests that it is the interplay between roll-streak cycles of different \modif{sizes} (wavelengths $L_z$ and $L_z/2$) that leads to longer lifetimes in the present model. This will be investigated in further detail in \S~\ref{sec:scaleinteraction}. \modif{Figure \ref{fig:waleffelifetimes}(b) also includes turbulence lifetimes when modes 7 or 9, two other structures absent from earlier ROMs, are neglected from the present model. Again, order-of-magnitude reductions of turbulence lifetimes are obtained when such modes are neglected, which highlights that all modes of the reduced basis are important in the chaotic dynamics. Neglecting mode 7 leads to turbulence lifetimes nearly identical to the ones obtained when mode 12 ($L_z/2$ rolls) is removed from the model, which suggests an important relationship between these structures. Further analysis of the role of modes 7 and 9 is postponed to section \ref{sec:scaleinteraction}.}

\begin{figure}
\begin{subfigure}{0.49\textwidth}
\includegraphics[width=1.0\textwidth]{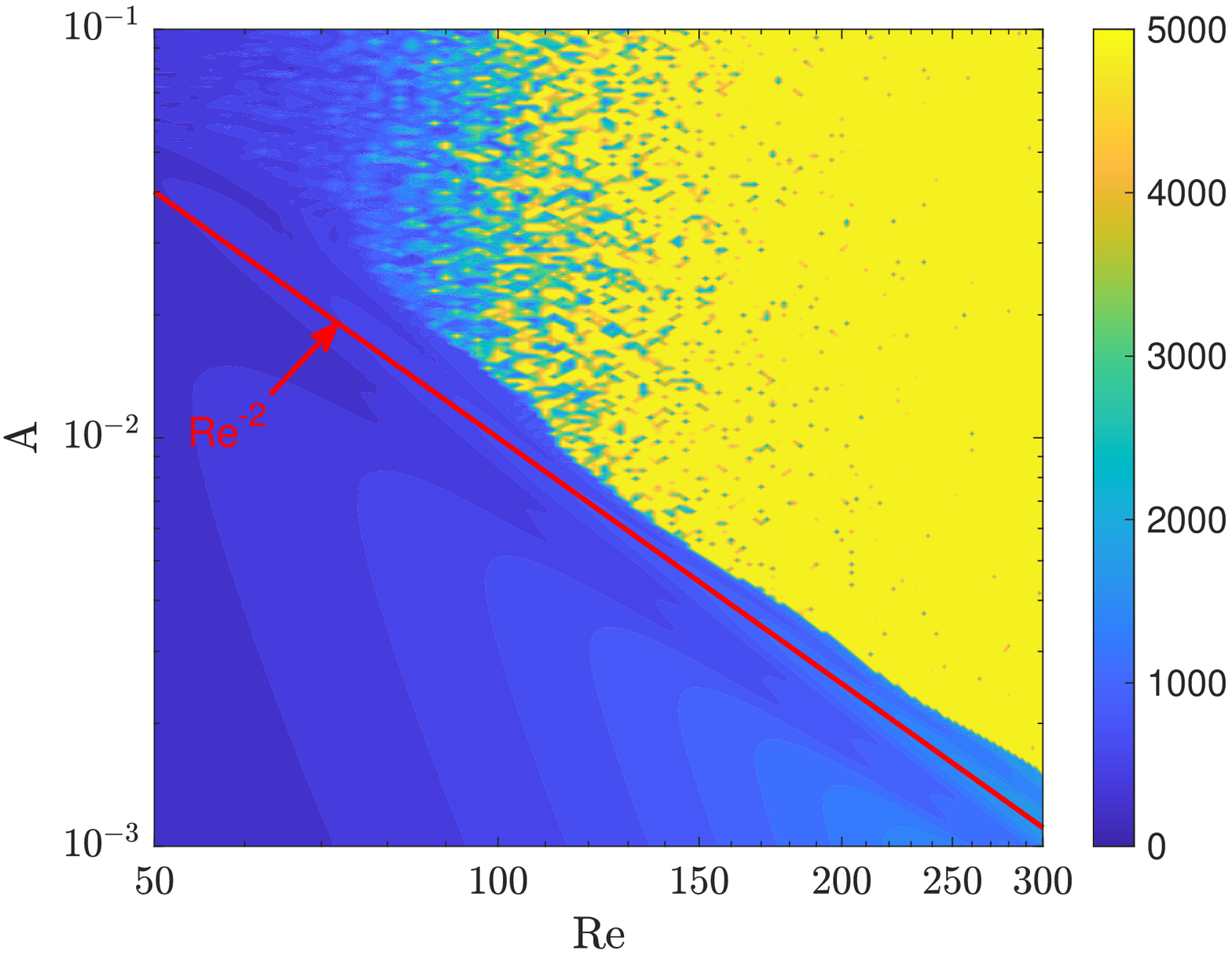}
\caption{Full model}
\end{subfigure}
\begin{subfigure}{0.49\textwidth}
\includegraphics[width=1.0\textwidth]{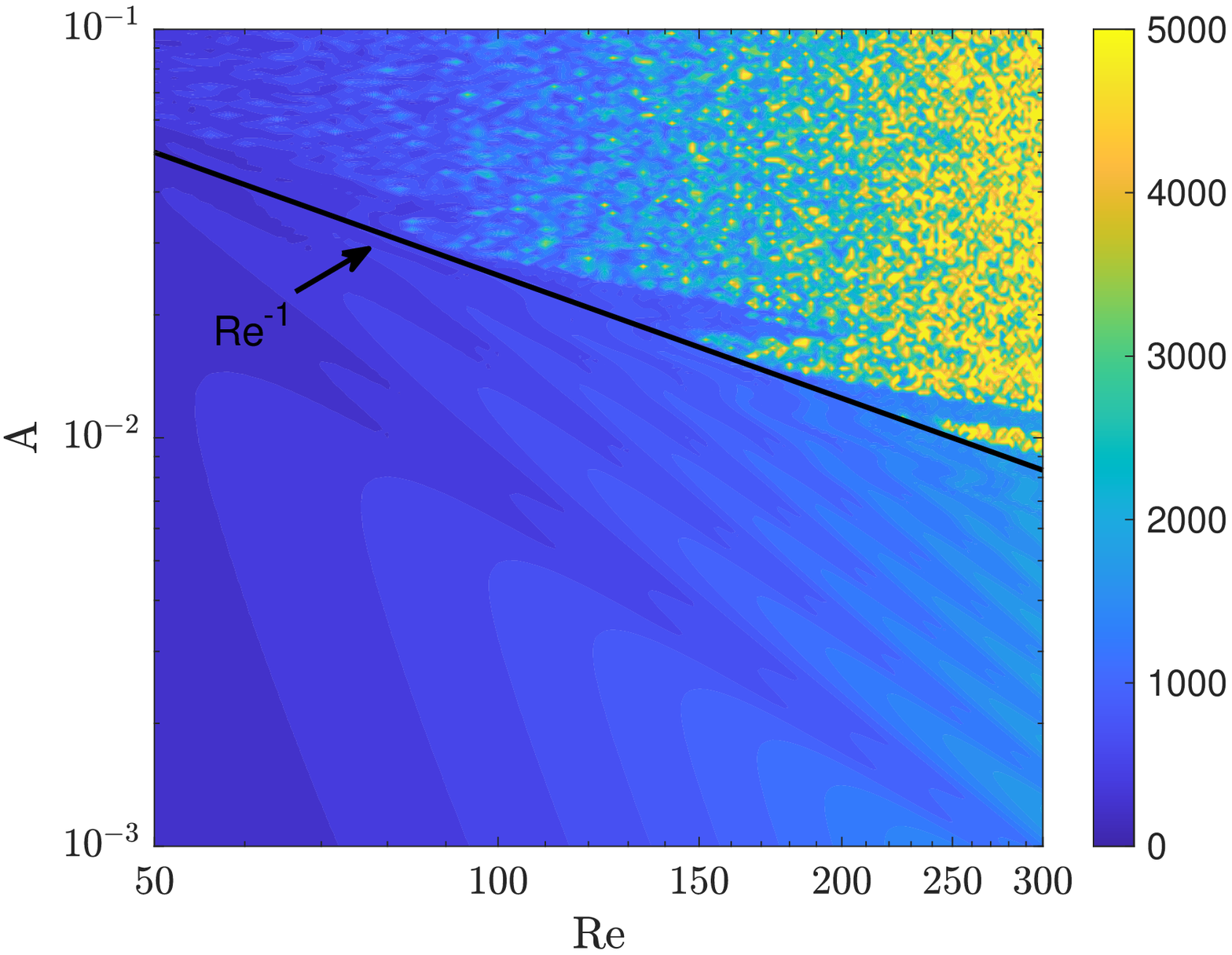}
\caption{Mode 12 neglected}
\end{subfigure}
\caption{Turbulence lifetime of the Waleffe-flow model following an initial disturbance given by $(a_1-1)=a_2=...=a_{12}=A$. Simulations carried out until $t=5000$.}
\label{fig:amp_lifetime_waleffe}
\end{figure}

As the system is linearly stable, transition to turbulence is related to finite-amplitude disturbances. This is investigated by running simulations with initial condition given by $a_1=1+A$ and $a_2=a_3=...=a_{12}=A$ and tracking lifetimes of turbulent behaviour. Results of such simulations, ran for 5000 convective time units, are shown in figure \ref{fig:amp_lifetime_waleffe}(a). The plot show features of a chaotic saddle, with small changes in the initial disturbance leading to significantly different lifetimes, similar to what is seen in the models by \citet{eckhardt1999transition} and \citet{moehlis2004low}. However, compared to the aforementioned works, the present model displays a higher ``density'' of initial conditions that reach long turbulence lifetimes (the predominantly yellow region for $Re>100$), again indicating that the present model has turbulence-maintaining features that are absent from the cited models. This can be more clearly seen by the analysis of the results of the present model with mode 12 neglected, shown in figure \ref{fig:amp_lifetime_waleffe}(b). Similar to the observations in the MFE model, a large variation of lifetimes is seen for sufficiently high disturbance amplitude, which can be seen from the grained aspect of the green and yellow region in the figure. Such wide distribution of lifetimes does not occur in the full model, where it becomes extremely unlikely to have relaminarisations after brief transients for higher $Re$.

For low disturbance amplitudes a minimal threshold for turbulence is observed, scaling approximately with $Re^{-2}$. A $Re^{-1}$ amplitude scaling was reported by \citet{eckhardt1999transition}, and such scaling is matched by the present model if mode 12 is neglected, shown in figure \ref{fig:amp_lifetime_waleffe}(b). The present $Re^{-2}$ amplitude threshold depends on the specific choice of disturbances introduced to the system, which were here taken to be deviations from the laminar solution with constant amplitude for all modes, but comparison of figures \ref{fig:amp_lifetime_waleffe}(a) and (b) shows that the inclusion of the two \modif{wavelengths}, $L_z$ and $L_z/2$, drastically changes the transient behaviour, with lower-amplitude disturbances that are capable of inducing transition. These observations are reminiscent of the findings of \citet{dawes2011turbulent}, who considered a Galerkin model of Waleffe flow with the 8 modes of Waleffe in streamwise and wall-normal directions, but with a large number of spanwise Fourier modes. This leads to a chaotic saddle with high ``density'' of long lifetimes, similar to the one of figure \ref{fig:amp_lifetime_waleffe}(a), with a minimal amplitude threshold for transition scaling with $Re^{-2.1}$ and $Re^{-2.3}$ depending on the choice of initial conditions.


\subsection{Couette flow}

\subsubsection{Turbulence lifetimes}

\begin{figure}
\begin{subfigure}{0.49\textwidth}
\includegraphics[width=1.0\textwidth]{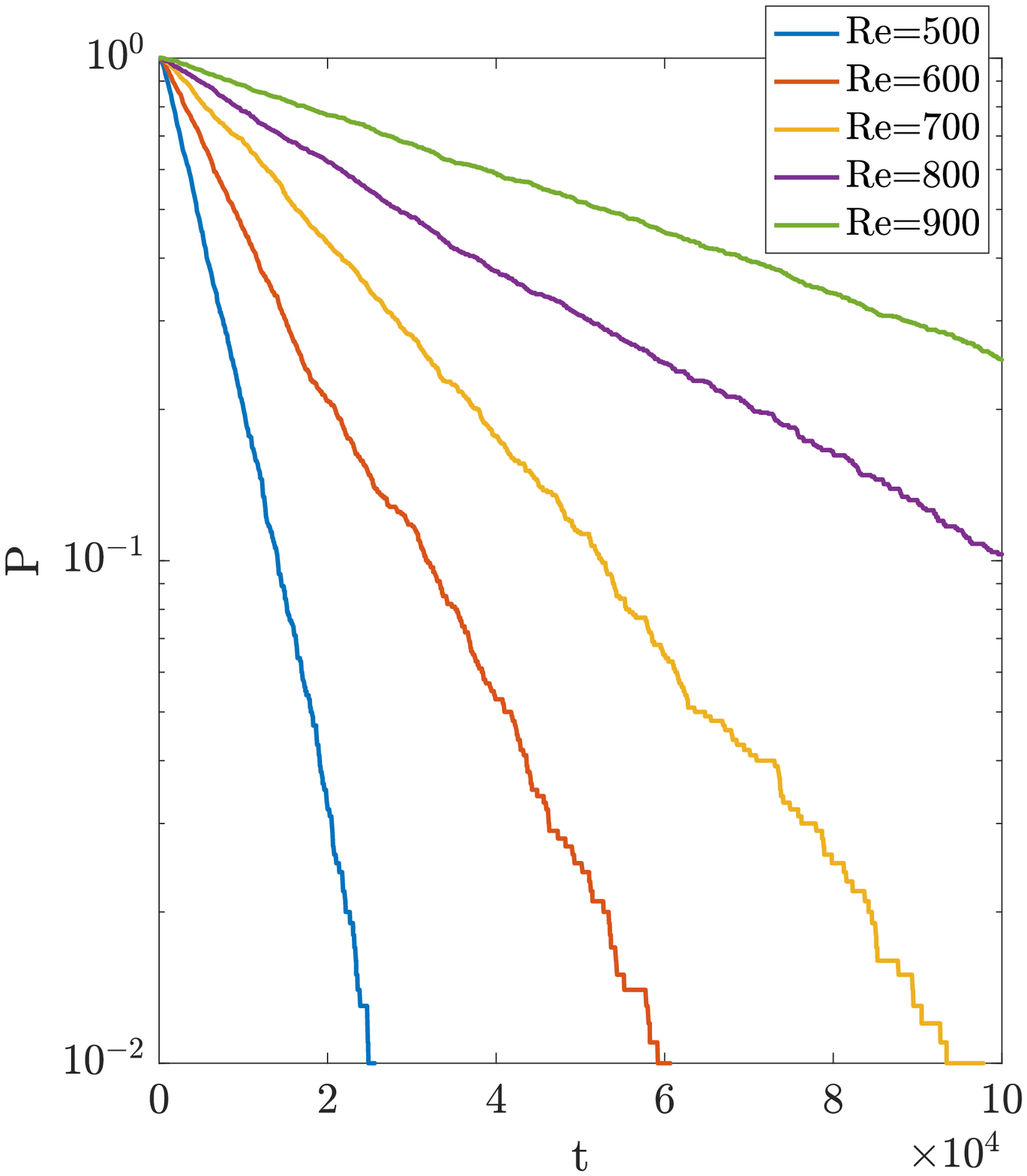}
\caption{Probability of chaotic behaviour until time $t$, $P(t)$}
\end{subfigure}
\begin{subfigure}{0.49\textwidth}
\includegraphics[width=1.0\textwidth]{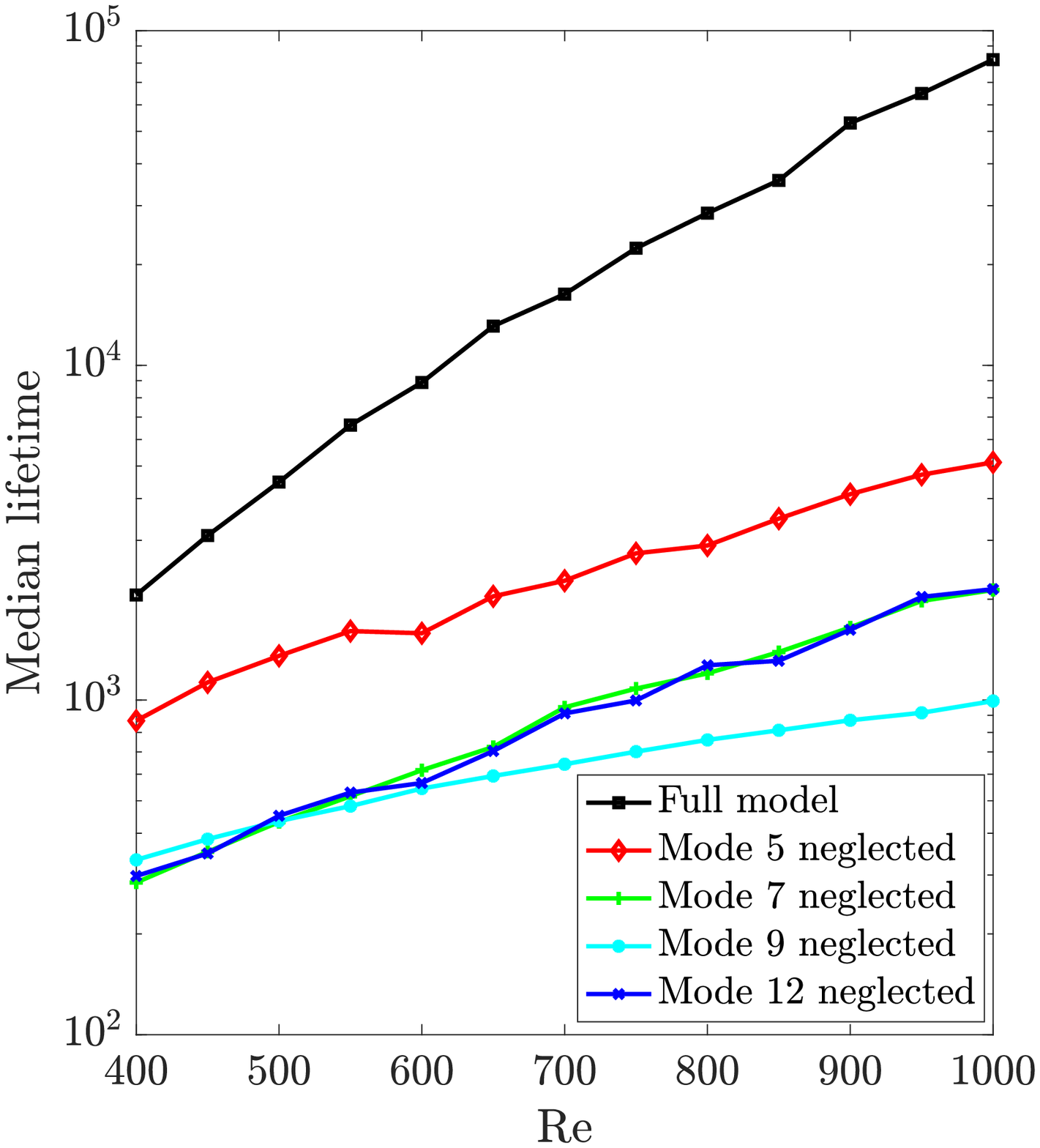}
\caption{Median lifetime as a function of Reynolds number}
\end{subfigure}
\caption{Couette-flow model lifetime statistics, taken from 1000 simulations with random initial conditions with norm equal to 0.3.}
\label{fig:couettelifetimes}
\end{figure}

We maintain for Couette flow a computational domain with $L_x=4\pi$ and $L_z=2\pi$. Results for other domain sizes are shown in Appendix~\ref{sec:boxsize}, showing that the results in this section are not due to this specific choice of domain. Sample time series of the model for Couette flow display an overall behaviour similar to figure \ref{fig:waleffetimeseries}, with a chaotic transient that settles back to the laminar solution after a long lifetime, \modif{significantly larger than the typical ``bursting period'' of Couette flow, which is about 100 convective time units for $\mathrm{Re}=400$~\cite{hamilton1995regeneration}.} Following the procedure for Waleffe flow in the preceding section, we have run several simulations of the Couette-flow model in order to obtain the probability $P(t)$ of turbulent flow after time $t$. Results are shown in figure~\ref{fig:couettelifetimes}, and display similar features of figure \ref{fig:waleffelifetimes}, with exponential decay of $P(t)$ with increasing $t$, at a slower rate for larger Reynolds numbers, leading to a fast increase of the median lifetime with $Re$. However, for Couette flow the median lifetimes are lower than what is observed for Waleffe flow. This may be due to the stronger constraints to the fluctuations in Couette flow, which should be strictly zero on the walls. \citet{kreilos2014increasing} report a median lifetime of about 400 for Couette flow at $Re=400$ in a computational box ($L_x=2\pi, L_z=\pi$); \andre{for this box size and Reynolds number, the results in figure \ref{fig:boxsize} in the Appendix show a median lifetime equal to 220. Keeping in mind that an exact match with a full simulation is not expected given the low number of degrees of freedom in the ROM, the present results indicate} that turbulence lifetimes in the present model are consistent with what is observed in numerical simulations, despite the severe truncation to 12 degrees of freedom. As was observed for Waleffe flow in figure \ref{fig:waleffelifetimes}(b), neglect of modes 5 ($L_z$ rolls) or 12 ($L_z/2$ rolls) leads to significantly lower turbulence lifetimes, indicating that both wavelengths are relevant for the dynamics in the model.

\modif{Figure~\ref{fig:couettelifetimes} also includes turbulence lifetimes for the model discarding either mode 7 (wall-normal vortices 2) or mode 9 (oblique wave 2), which are absent from earlier models, as discussed in section \ref{sec:strategy}. As for Waleffe flow, neglecting either mode also leads to a reduction of more than an order of magnitude of lifetimes. Removal of mode 7 from the model leads to the same lifetimes obtained when mode 12 is neglected, highlighting that both structures are dynamically related. Further discussion on this is presented in section \ref{sec:scaleinteraction}. The observed reductions of turbulence lifetimes once a mode is removed from the model show that all structures represented with the present basis are important in maintaining chaotic motion. The new modes in the present system are thus worthy of further study to explore their role in turbulence dynamics.}

\begin{figure}
\centerline{\includegraphics[width=0.8\textwidth]{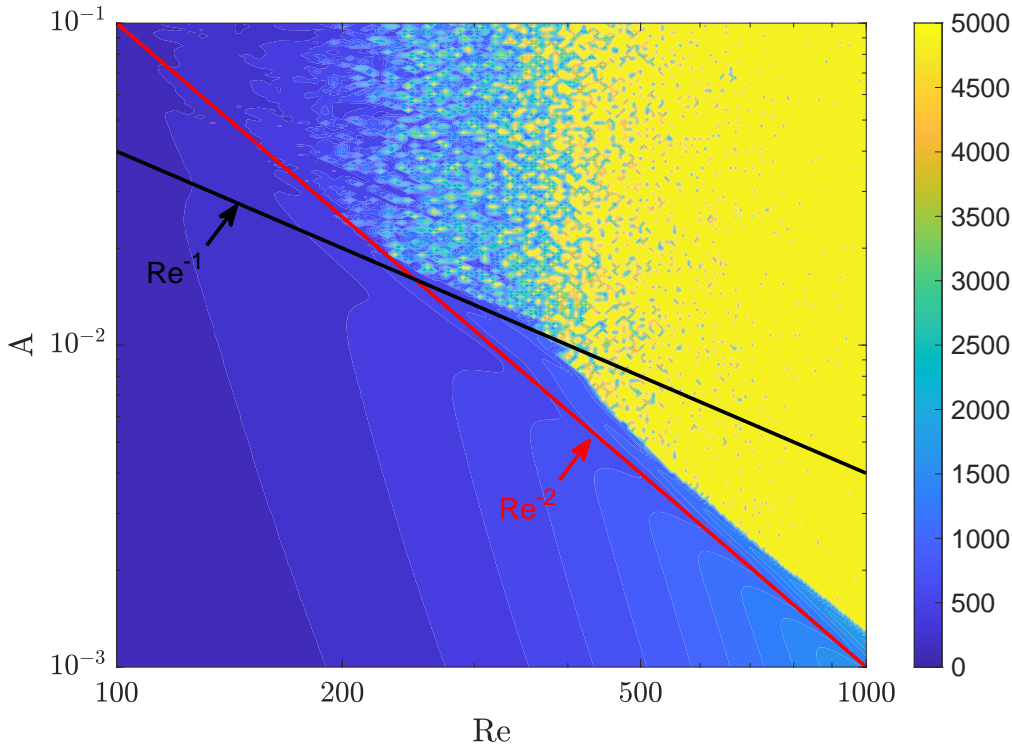}}
\caption{Turbulence lifetime of the Couette-flow model following an initial disturbance given by $a_1=a_2=...=a_{12}=A$. Simulations carried out until $t=5000$.}
\label{fig:amp_lifetime_couette}
\end{figure}

The impact of initial disturbance amplitude on turbulence lifetime is studied in figure \ref{fig:amp_lifetime_couette}, with simulations carried out up to $t=5000$ for a disturbance given by $a_1=a_2=...=a_{12}=A$. The results show again features of a chaotic saddle, similar to what was observed for Waleffe flow in figure \ref{fig:amp_lifetime_waleffe}. For $Re$ between 200 and 400 this threshold scales approximately with $Re^{-1}$, whereas higher $Re$ see a transition threshold scaling with approximately $Re^{-2}$, the same Reynolds number trend of the Waleffe-flow model. As discussed in the Introduction, a number of studies have shown that wall-bounded flows have a transition due to disturbances of finite amplitude, whose minimal value for transition scales with $Re^{-\gamma}$. The present value of $\gamma=2$ is of course severely restrained by the truncation of the system to 12 modes. \citet{duguet2013minimal} have found minimal-amplitude disturbances for transition in Couette flow with amplitude scaling of $\gamma=1.35$, a value significantly lower than the scaling found here.

Bearing such difference in mind, the fact that we obtain $\gamma>1$ indicates that low-amplitude disturbances are able to exploit non-linear mechanisms in the flow leading to transition. \modif{Transient growth of streaks from streamwise vortices has an amplitude gain that scales with $\mathrm{Re}$~\cite{trefethen1993hydrodynamic}, which alone would lead to $\gamma=1$; there are thus other mechanisms at play.} Non-linear mechanisms only redistribute energy and do not lead to growth~\citep{henningson1996comment}, but such redistribution may exploit linear mechanisms, as discussed by \citet{trefethen1993hydrodynamic} and \citet{baggett1995mostly}. Here, the $\gamma=2$ scaling appears for $Re>400$ and is coincident with the emergence of a chaotic saddle with higher ``density'' of longer lifetimes in figure \ref{fig:amp_lifetime_couette}, which again shows that the present model has intrinsic dynamics that help maintain turbulence for longer lifetimes. 

\subsubsection{Comparison with direct numerical simulations}

Differently from Waleffe flow, there are many available experimental and numerical results for plane Couette flow, which may be used to verify if the model predictions agree with the expected statistics. \andre{This is a fundamental difference between the present model and earlier ROMs based on Waleffe flow~\cite{waleffe1997self,moehlis2004low}, whose statistics could not be compared to reference numerical data. Given the low order of the system, close quantitative matches are not expected, as a reproduction of the turbulence physics would require a resolution similar to direct numerical simulation (DNS). However, a ROM should recover at least some qualitative trends observed in the data to ensure that meaningful physics are retained in the truncated system.} 

The results of sufficiently long simulations of the Couette model may be compared to DNS results. Comparisons were performed with results from the ChannelFlow pseudo-spectral solver~\citep{gibson2019channelflow}. Simulations were run in the same numerical box of $L_x=4\pi$ and $L_z=2\pi$, also imposing the symmetry $u,v,w(-x,-y,-z)=u,v,w(x,y,z)$. 64 Fourier modes (96 if dealiasing is considered) were used in the simulation, and 65 Chebyshev polynomials were adopted for the discretisation in the wall-normal direction. For $Re=800$ the simulation leads to a friction Reynolds number equal to 55, and grid spacings of 11 wall units in streamwise and 5.5 wall units in spanwise direction, ensuring a resolution compatible with DNS. Simulations for $Re=500$ and 800 were carried out for 700 convective time units, discarding initial transients.

\begin{figure}
\begin{subfigure}{0.49\textwidth}
\includegraphics[width=1.0\textwidth]{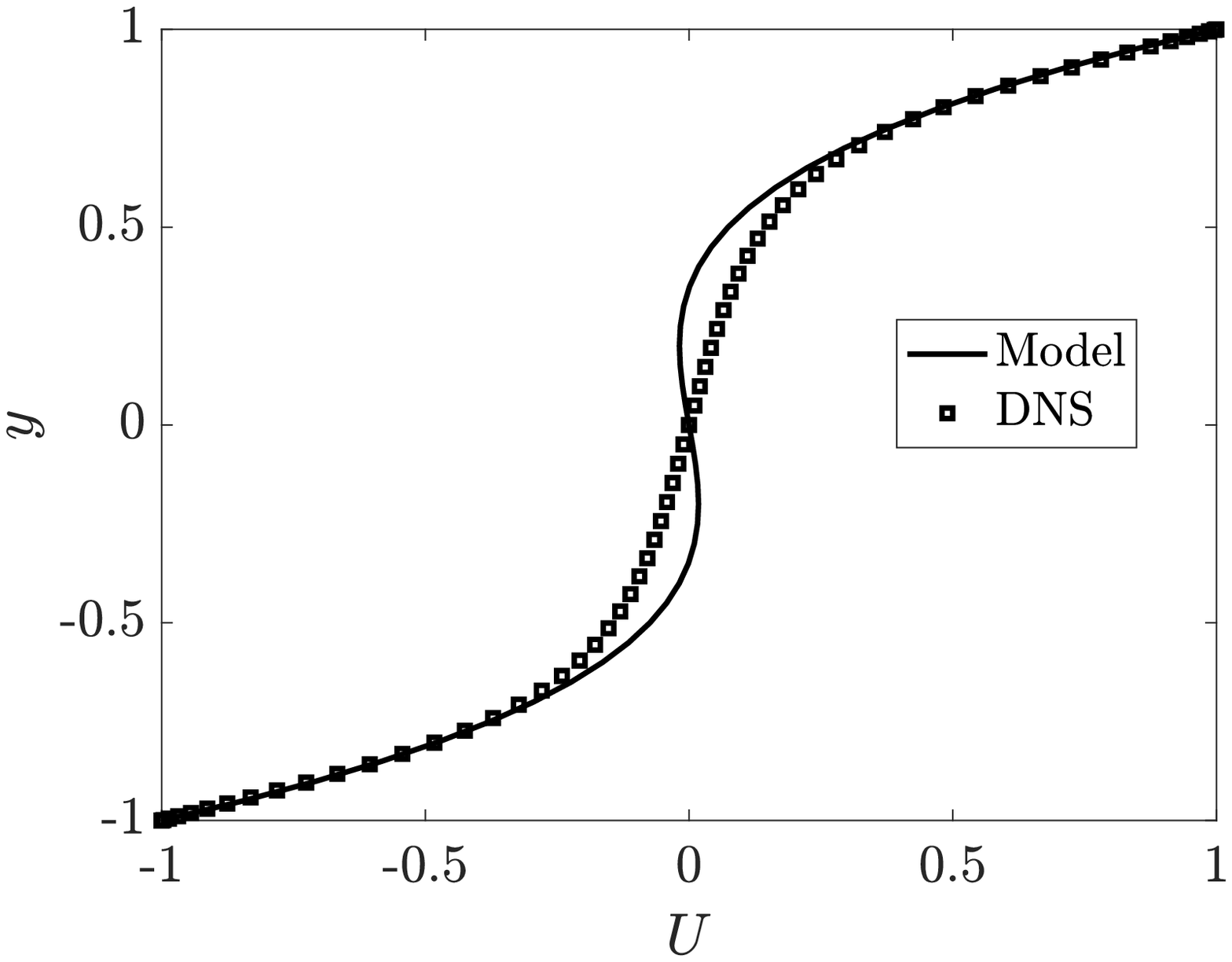}
\caption{Mean flow, $Re=500$}
\end{subfigure}
\begin{subfigure}{0.49\textwidth}
\includegraphics[width=1.0\textwidth]{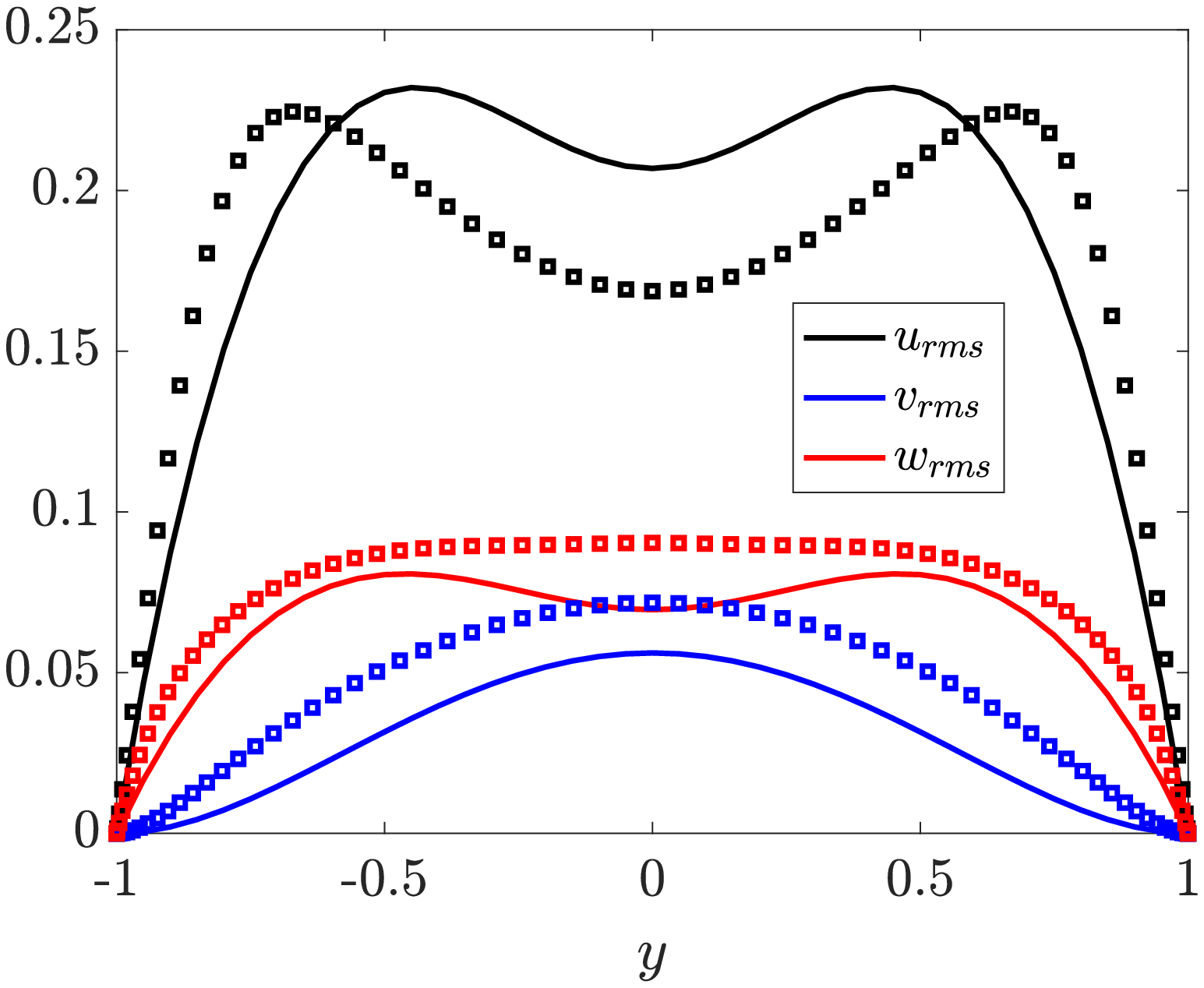}
\caption{RMS values, $Re=500$}
\end{subfigure}
\begin{subfigure}{0.49\textwidth}
\includegraphics[width=1.0\textwidth]{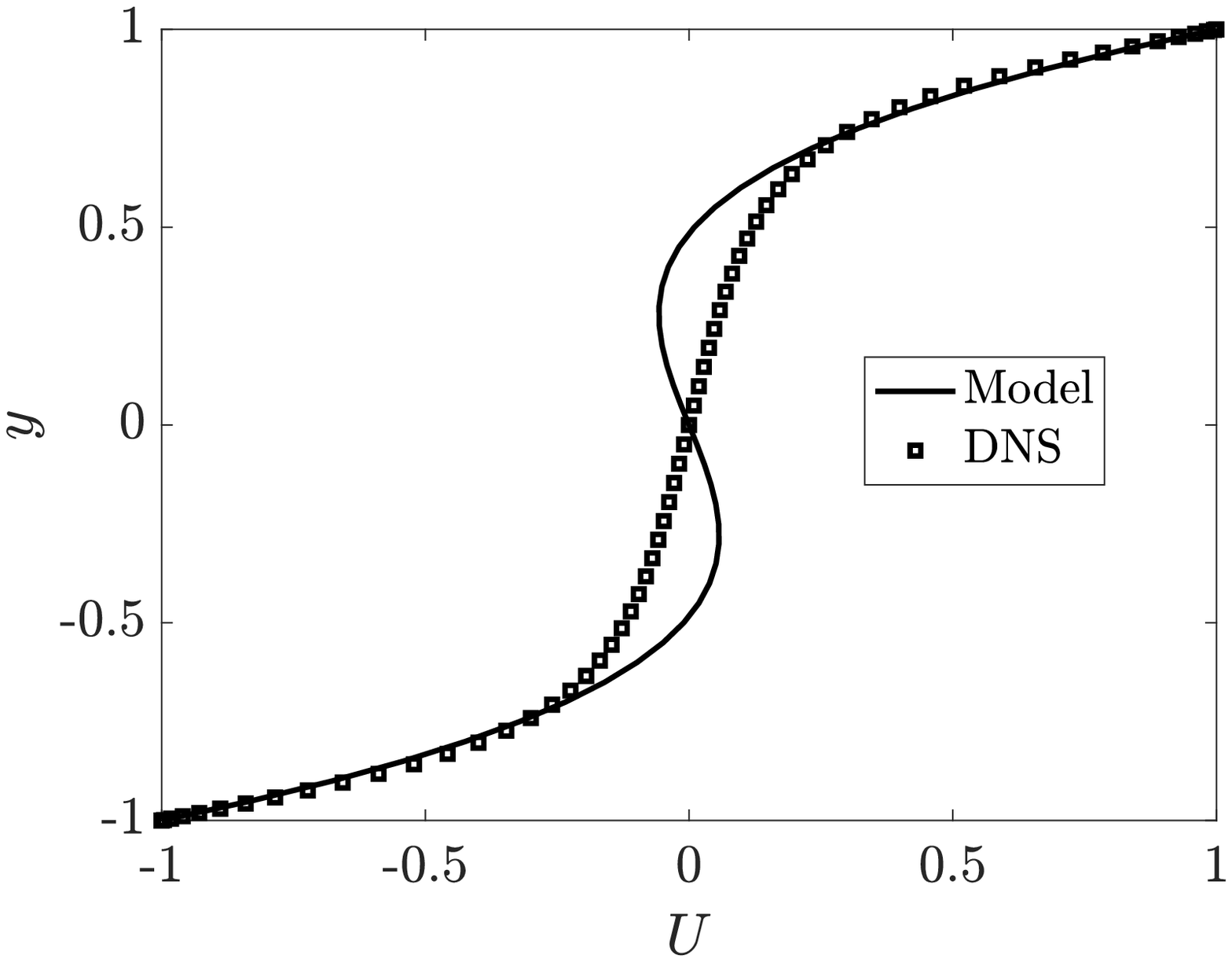}
\caption{Mean flow, $Re=800$}
\end{subfigure}
\begin{subfigure}{0.49\textwidth}
\includegraphics[width=1.0\textwidth]{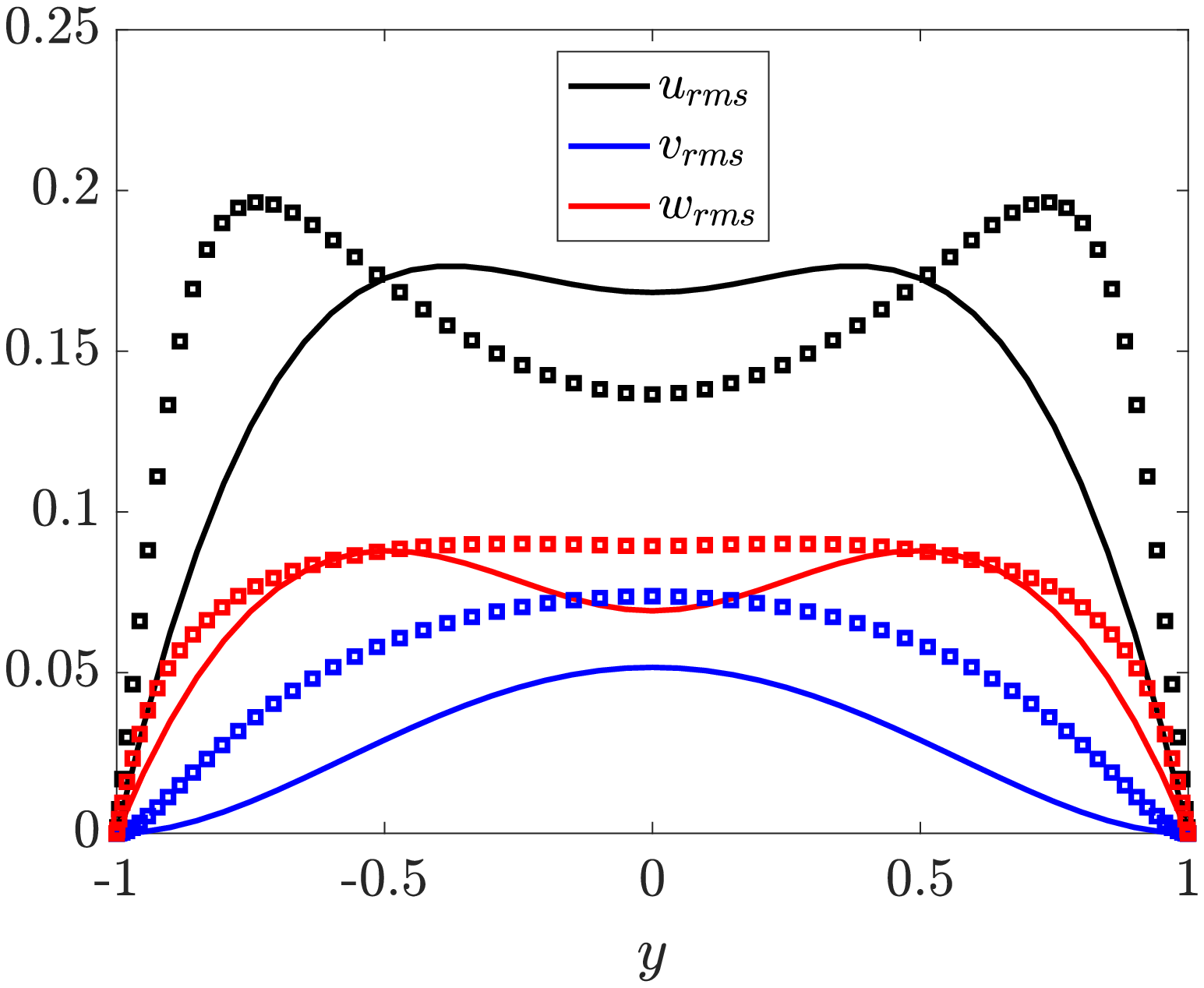}
\caption{RMS values, $Re=800$}
\end{subfigure}
\caption{Comparison between statistics of model (full lines) and DNS (symbols).}
\label{fig:DNSstats}
\end{figure}

The mean velocity profiles and RMS of velocity fluctuations from the model, taken from 5000 convective time units after initial transients, are compared to the DNS results in figure \ref{fig:DNSstats}, for $Re=500$ and 800. A reasonable agreement is seen between the model and the DNS statistics, particularly for $Re=500$. The ROM has only twelve degrees of freedom, and is thus unable to reproduce the details of all fluctuations in the DNS. The mean wall shear is nonetheless reproduced for both Reynolds numbers, but with errors in the mean flow in the central region likely due to its representation by a single mode, $\mathbf{u}_1$. \andre{Similar errors in the mean temperature profile are also observed for low-order Galerkin models of Rayleigh-B\'enard convection~\cite{saltzman1962finite}.} \modif{Despite the inflectional shape, tests with the present reduced-order model show that the mean flow does not present a linear instability.}

The overall shapes and peak values of the RMS of the three velocity components are reproduced by the model, particularly for $u$ and $w$, with a more visible mismatch in the RMS of $v$.  The comparison between the RMS values is better for $Re=500$, which may be attributed to the lower range of turbulent scales in this low-Re flow. \andre{The smallest scales in wall-bounded turbulent flows are known to scale with viscous units, and increasing the Reynolds number leads to smaller near-wall structures that cannot be represented with the set of twelve modes.} The reduction of RMS values of $u$ seen as the Reynolds number is increased from 500 to 800 is also obtained for the model. \andre{Notice that the errors in the RMS profiles are lower for the present model than in the POD-Galerkin models by \citet{smith2005low}, who report differences in peak values of about 50\%, even though the mean flow in their formulation matches the DNS by construction.}

\begin{figure}
\centerline{\includegraphics[width=0.48\textwidth]{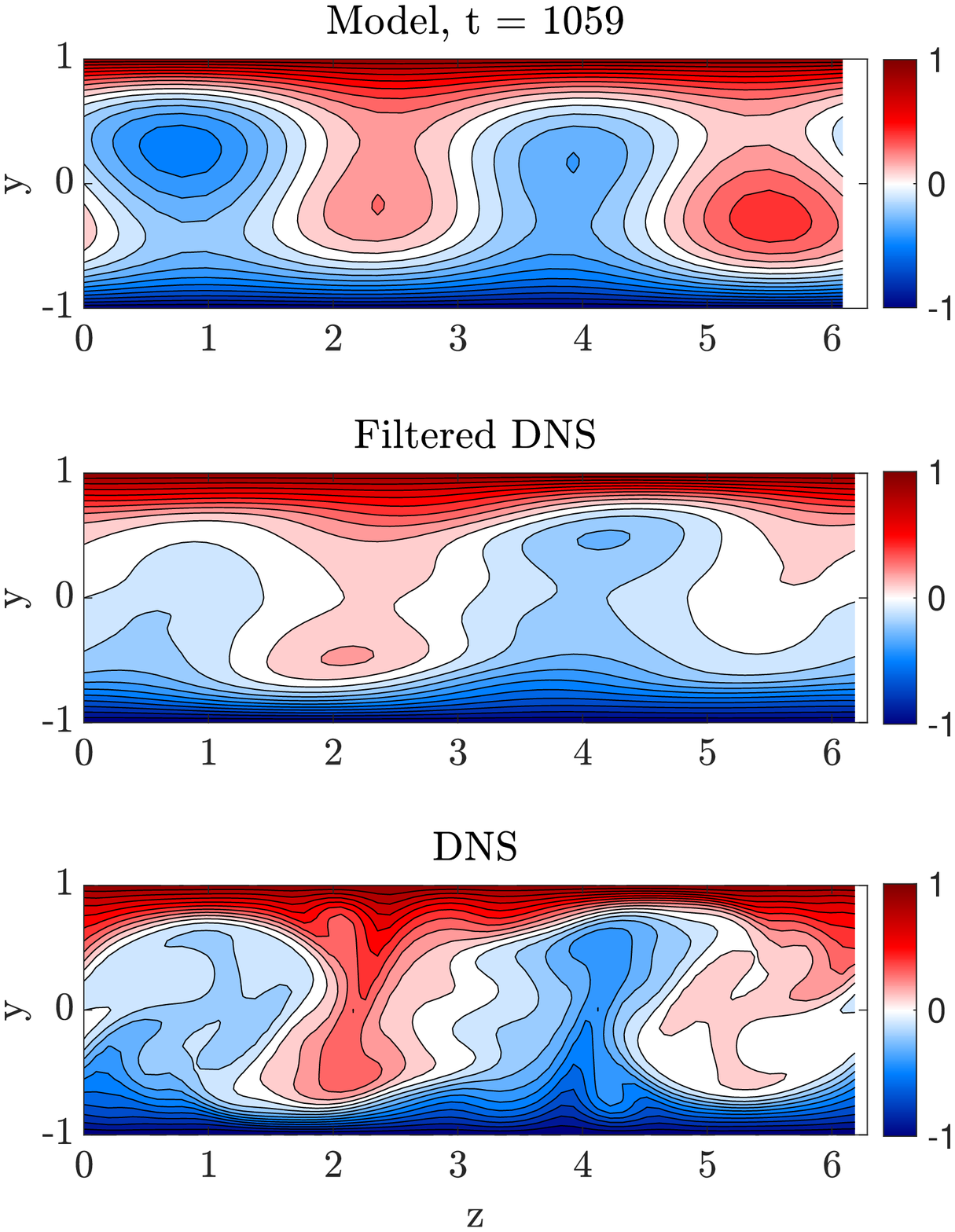}\includegraphics[width=0.48\textwidth]{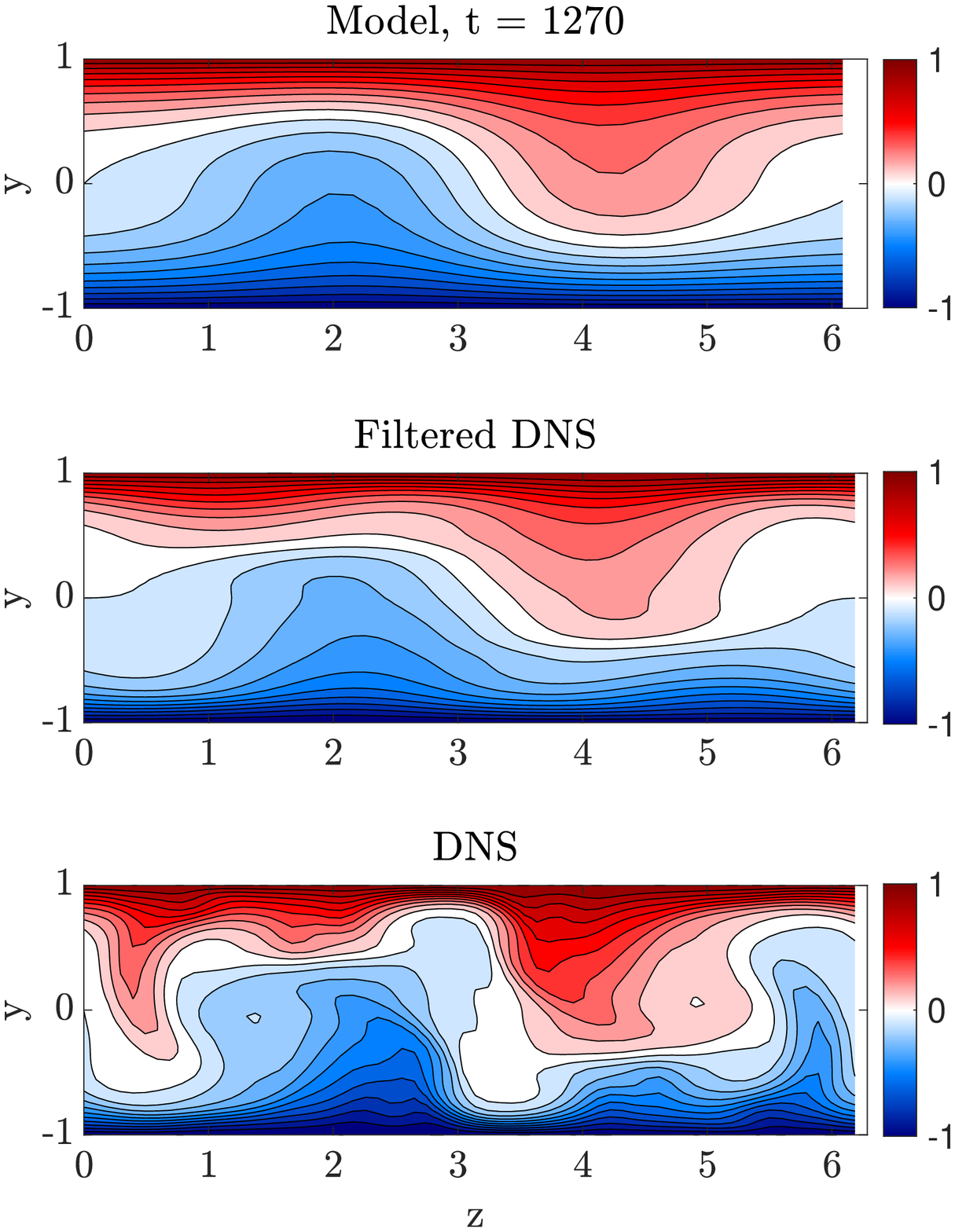}}
\caption{Sample cross section of the flow as predicted by the model (top row) and extracted from filtered (middle row) and full DNS (bottom row). Colours show the instantaneous streamwise velocity $u$. Left and right columns refer to two different sample timesteps. See the Supplemental Material for an animated version of this figure\cite{prffootnote}.}
\label{fig:DNSsnapshot}
\end{figure}

A comparison between cross-sections of sample snapshots from the model and the DNS for $Re=500$ is seen in figure \ref{fig:DNSsnapshot}. The selection of snapshots is arbitrary, but we notice that other times for both model and DNS display the same overall behaviour, with the presence of streaks with varying amplitude; we have selected two fields that display similar features for comparison. For the snapshots portrayed in figure \ref{fig:DNSsnapshot}, we notice that the main overall features in the DNS are also present in the model, with \modif{the snapshot in figure \ref{fig:DNSsnapshot}(a) displaying two pairs of positive and negative streaks (i.e. a dominance of mode 11), whereas figure \ref{fig:DNSsnapshot}(b) portrays a time with dominance of a single pair of streaks (mode 4).} Similar structures appear in the DNS and in the model, although the DNS has a much broader range of spatial scales, as expected, especially near the walls. \andre{To show which structures in the DNS may be represented in the model, we have filtered the DNS field so as to retain spanwise wavenumbers equal to 0, $\pm \gamma$ and $\pm 2\gamma$; the resulting field is labelled as ``Filtered DNS'' in figure \ref{fig:DNSsnapshot}, with structures that resemble more closely the ROM result. The video in the supplemental material shows a time series of the ROM, filtered and full DNS fields\cite{prffootnote}. The instantaneous structures are of course different, but the fields display similar motions, confirming that the observed agreement is not fortuitous.}

\section{The role of structure interactions}
\label{sec:scaleinteraction}

We now attempt to explore how $L_z$ and $L_z/2$ streaks and rolls interact in the dynamical system. For both Waleffe and Couette flows the presence of both $L_z$ and $L_z/2$ rolls was seen to be relevant to maintain longer turbulence lifetimes, as figures \ref{fig:waleffelifetimes}(b) and \ref{fig:couettelifetimes}(b) show that the removal of any of these modes leads to reductions of more than an order of magnitude in the median lifetime.  Instead of simply removing a mode from the system, we track more closely how the energy exchanges in the system couple the two wavelengths. This can be done by computing the energy budget, in a procedure analogous to \citet{noack2005need}, but here used for fluctuations around the laminar solution. For a given mode $a_i$, multiplication of its equation by $a_i$ shows that the energy varies according to
\begin{equation}
\frac{\mathrm{d} (a_i^2 / 2)}{\mathrm{d} t}
=
\sum_j{L_{i,j}a_i a_j}
+
\sum_j{\sum_k{{Q_{i,j,k}}a_i a_j a_k}}
+
F_i a_i.
\end{equation}
Averaging over long times with chaotic dynamics leads to
\begin{equation}
\sum_j{L_{i,j}\overline{a_i a_j}}
+
\sum_j{\sum_k{{Q_{i,j,k}}\overline{a_i a_j a_k}}}
+
F_i \overline{a_i}
= 0 
\end{equation}
where the overbar denotes time averaging. This allows an evaluation of the averaged energy transfer induced by each term of the Galerkin system. \modif{Here we will focus on the model for Couette flow, as this setup is more studied in the literature; however, a similar analysis was carried out for the Waleffe-flow model with very similar results, which will not be shown here for conciseness.}  For the equations in fluctuation form, as in the Couette flow model, the linear term has a viscous component that is dissipative, and another term that represents coupling with the laminar solution. The quadratic terms are conservative: a given mode gains energy that is extracted from another mode. Finally, the forcing term is zero in the Couette model.

For Couette flow, the equation for the $L_z$ streaks, mode 4, is
\begin{eqnarray}
\frac{\mathrm{d} a_4}{\mathrm{d} t}=
\underbrace{-\frac{a_{4}\,\left(\gamma ^2+\frac{5}{2}\right)}{\mathrm{Re}}}_{I}
\underbrace{-\frac{3\,\sqrt{21}\,a_{5}\,\gamma }{14\,k_{\beta,\gamma}}-\frac{3\,\sqrt{10}\,a_{1}\,a_{5}\,\gamma }{4\,k_{\beta,\gamma}}}_{II}
\underbrace{-\frac{3\,\sqrt{30}\,a_{2}\,a_{6}\,\gamma ^2}{14\,k_{\alpha,\gamma}}}_{III}
\nonumber \\
\underbrace{-\frac{\sqrt{10}\,a_{3}\,a_{8}\,\gamma }{2\,k_{\alpha,\beta}}}_{IV}
+\underbrace{\frac{\sqrt{30}\,a_{3}\,a_{10}\,\alpha \,\gamma ^2}{6\,k_{\alpha,\beta}\,k_{\alpha,\beta,\gamma}}}_{V}
\end{eqnarray}
Multiplication of this equation by $a_4$ leads to
\begin{eqnarray}
\frac{\mathrm{d} (a_4^2 / 2)}{\mathrm{d} t}=
\underbrace{-\frac{a_{4}^2\,\left(\gamma ^2+\frac{5}{2}\right)}{\mathrm{Re}}}_{I}
\underbrace{-\frac{3\,\sqrt{21}a_4 a_{5}\,\gamma }{14\,k_{\beta,\gamma}}-\frac{3\,\sqrt{10}a_4\,a_{1}\,a_{5}\,\gamma }{4\,k_{\beta,\gamma}}}_{II}
\underbrace{-\frac{3\,\sqrt{30}a_4\,a_{2}\,a_{6}\,\gamma ^2}{14\,k_{\alpha,\gamma}}}_{III}
\nonumber \\
\underbrace{-\frac{\sqrt{10}\,a_4\,a_{3}\,a_{8}\,\gamma }{2\,k_{\alpha,\beta}}}_{IV}
+\underbrace{\frac{\sqrt{30}\,a_4\,a_{3}\,a_{10}\,\alpha \,\gamma ^2}{6\,k_{\alpha,\beta}\,k_{\alpha,\beta,\gamma}}}_{V}
\label{eq:energytermsa4}
\end{eqnarray}
which shows that the first term (marked as group $I$) is related to viscous dissipation, the second term is related to coupling with the laminar solution, and the remaining terms are non-linear interactions in the model. The second and third terms are gathered in group $II$, which is related to the lift-up effect. The second term allows extraction of energy from the laminar solution in the presence of rolls $a_5$, and the third term modifies the lift-up process due to mean-flow distortion $a_1$. Groups $III$, $IV$ and $V$ are non-linear interactions with various other modes, with group $III$ related to streak instability by \citet{waleffe1997self}.

\begin{figure}
\begin{subfigure}{0.49\textwidth}
\includegraphics[width=1.0\textwidth]{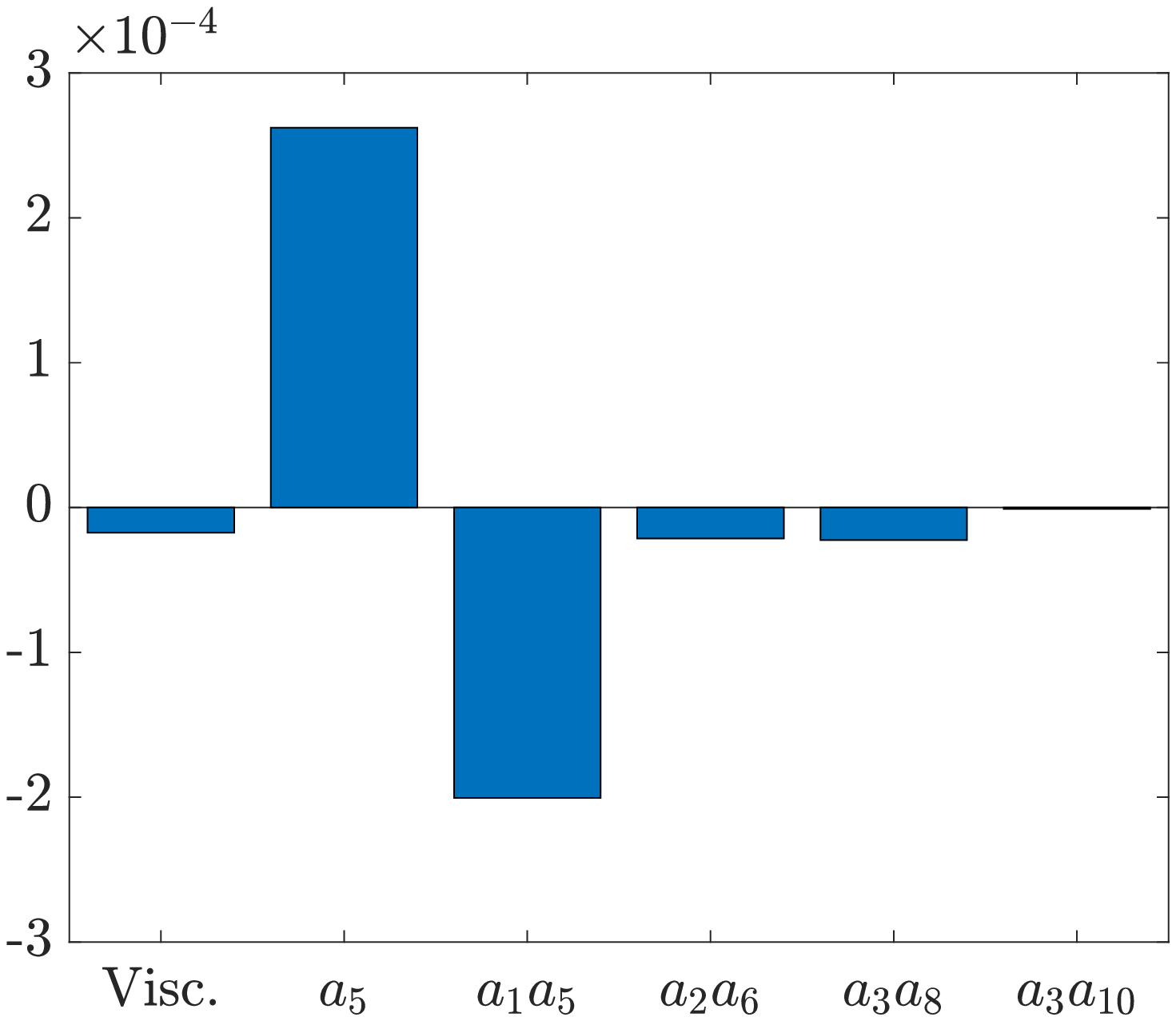}
\caption{Mode 4 ($L_z$ streak)}
\end{subfigure}
\begin{subfigure}{0.49\textwidth}
\includegraphics[width=1.0\textwidth]{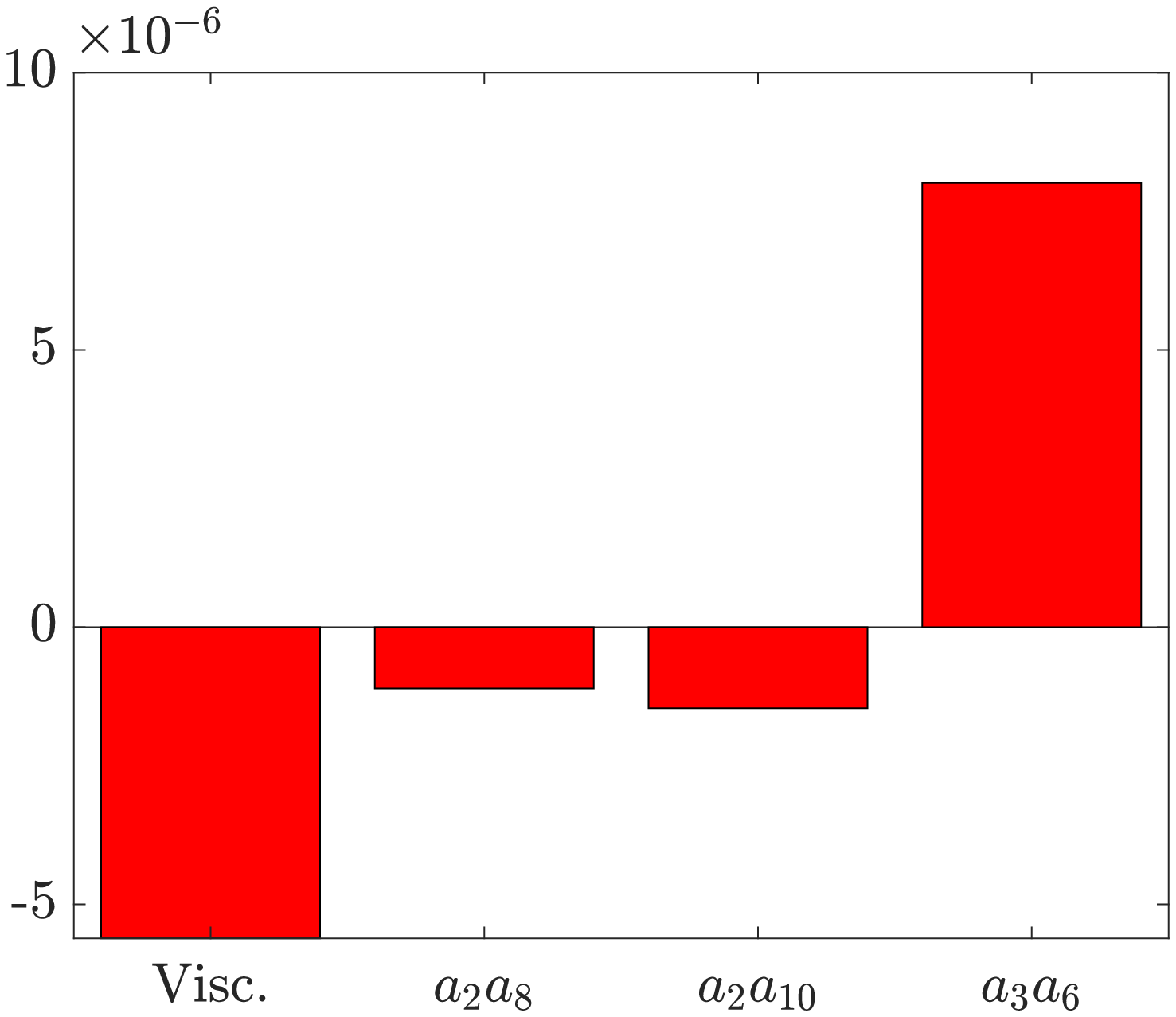}
\caption{Mode 5 ($L_z$ roll)}
\end{subfigure}
\begin{subfigure}{0.49\textwidth}
\includegraphics[width=1.0\textwidth]{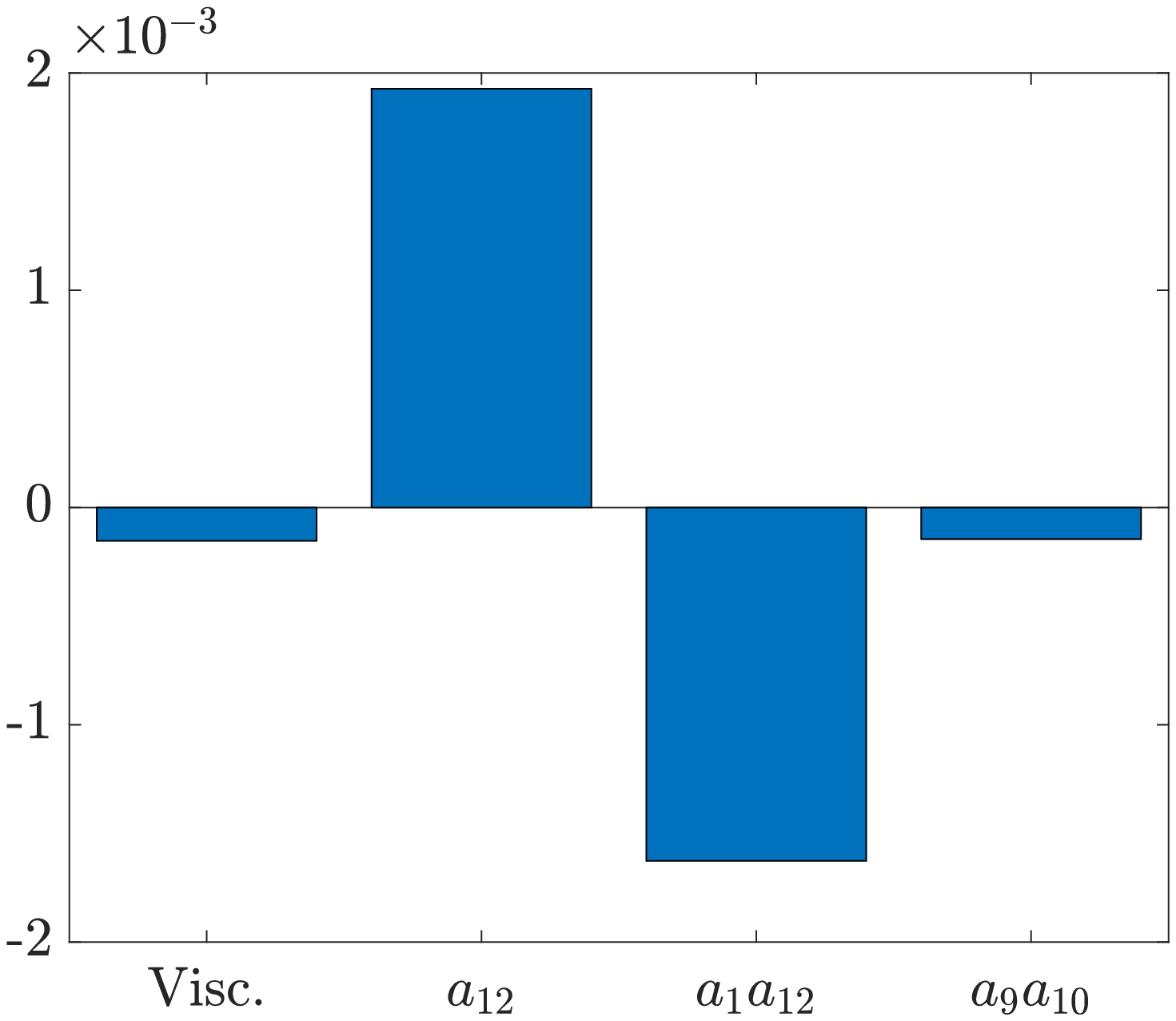}
\caption{Mode 11 ($L_z/2$ streak)}
\end{subfigure}
\begin{subfigure}{0.49\textwidth}
\includegraphics[width=1.0\textwidth]{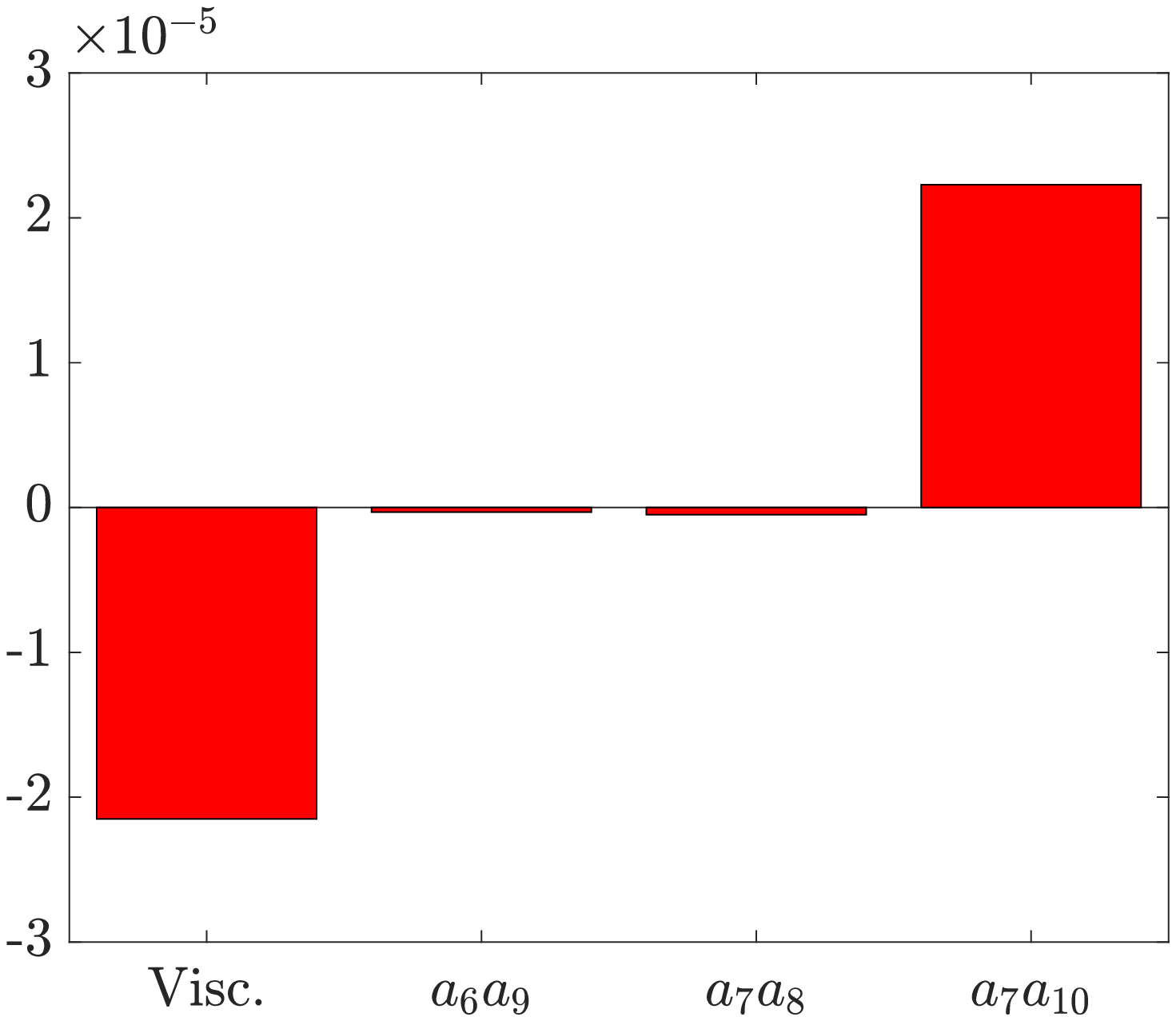}
\caption{Mode 12 ($L_z/2$ roll)}
\end{subfigure}
\caption{Energy budgets for $Re=500$. ``Visc.'' stands for losses related to the viscous term, whereas the other bars show energy contributions from each other linear and non-linear terms in the equations, here represented by the mode coefficients in each term.}
\label{fig:energybudget}
\end{figure}

In what follows we refer to a given mode $i$ by its time coefficient $a_i$ for convenience, to simplify notation since various non-linear terms will be examined. The energy budgets for modes $a_4$ ($L_z$ streak), $a_5$ ($L_z$ roll), $a_{11}$ ($L_z/2$ streak) and $a_{12}$ ($L_z/2$ roll) are shown in figure \ref{fig:energybudget}. Budgets were computed by evaluating the linear and non-linear terms in the energy equation for each mode, taken from the final half (5000 convective time units) of a simulation for $\mathrm{Re}=500$ with 10000 convective time units without relaminarisation. All budgets are closed within less than $1\%$. Streaks and rolls are chosen due to their known relevance in wall-bounded turbulence~\citep{hamilton1995regeneration}. The $L_z$ ($a_4$ and $a_5$) and $L_z/2$ ($a_{11}$ and $a_{12}$) spanwise wavelengths are not directly related through non-linear terms, as one length is absent from the equations of the other. However, the non-linear interactions with other modes in the system couple these \modif{modes} in a subtle way, as will be seen by the analysis of the budgets.

We start by the analysis of the $L_z$ streak $a_4$. Its energy comes from the laminar solution via the $a_5$ linear term, related to the lift-up mechanism based on the laminar solution, as this is the only term with positive contribution to the energy. If $a_5$ and $a_1a_5$ terms are added, we still have a positive contribution from the lift-up term $II$ in eq. \ref{eq:energytermsa4}, which represents lift-up including mean-flow distortion. Besides viscous losses, the streak has significant energy transfer to modes $a_2$ and $a_6$ through the $a_2a_6$ term, and to mode $a_8$ through the $a_3a_8$ term; the modes that receive energy may be inferred from the model equations (\ref{eq:couettemodel}), which show that the $a_2a_6$ term matches the sum of corresponding terms in the $a_2$ and $a_6$ equations, whereas the $a_3a_8$ term matches a term in the $a_8$ equation. The contribution of the remaining term $a_3a_{10}$ is small, showing that on average the energy transfer related to it is negligible. The $L_z$ rolls get their energy from the $a_3a_6$ term, which implies, from the model equations, an energy transfer from modes $a_3$ and $a_6$.

If we now turn our attention to the $L_z/2$ modes $a_{11}$ and $a_{12}$, we notice that the $L_z/2$ streak also gets energy through the lift-up effect, with term $a_{12}$ related to the laminar solution and term $a_1a_{12}$ showing a change in lift-up due to mean-flow distortion. The $L_z/2$ streak loses energy to modes $a_9$ and $a_{10}$ through the $a_9a_{10}$ term. The $L_z/2$ roll $a_{12}$ receives energy from modes $a_7$ and $a_{10}$ through the $a_7a_{10}$ term. \modif{This last observation provides an explanation for the same turbulence lifetimes obtained for the model with either mode $a_{12}$ or mode $a_7$ neglected, as seen in figures \ref{fig:waleffelifetimes}(b) and \ref{fig:couettelifetimes}(b); neglecting mode $a_7$ amounts to discarding the energy transfer towards mode $a_{12}$, such that the latter mode is not excited. Moreover, discarding mode $a_9$ may also be related to this process, as the equation for $a_7$ (\ref{eq:a7couette}) has a linear term with $a_9$. This term leads to a growth of energy of $a_7$, with energy extracted from the mean flow (see Reynold-stress term $a_7a_9$ in the mean-flow equation (\ref{eq:a1couette})). Thus, removal of mode $a_9$ from the model eliminates such linear mechanism, which in turn reduces the energy transferred to mode $a_7$ and thus weakens the forcing of the $L_z/2$ rolls $a_{12}$.}

The observation of the budgets in figure \ref{fig:energybudget} gives the impression that the $L_z$ and $L_z/2$ modes are uncoupled, as the bulk of energy transfers from one \modif{wavelength} is not directly related to the other. However, they are coupled to each other by the mean-flow mode $a_1$, which modifies the lift-up effect for both \modif{wavelengths}. There are also couplings through the other equations in the dynamical system, in a process that may be rather subtle. For instance, we have observed that mode $a_6$ mediates the energy transfer to the $L_z$ roll $a_5$, and mode $a_7$ gives energy to the $L_z/2$ roll $a_{12}$. As shown in table~\ref{tab:modes} (also in eq. (\ref{eq:couettemodes}) in appendix \ref{sec:equations}) , modes $a_6$ and $a_7$ are both wall-normal vortices with the same spatial shape, but phase-shifted by $\pi/2$ in streamwise and spanwise directions. Inspection of the equations for $a_6$ and $a_7$ in eq. (\ref{eq:couettemodel}) shows that these modes are coupled: there is an $a_7a_{11}$ term in the equation for $a_6$, and an $a_6a_{11}$ term in the equation for $a_7$. In terms of energy of modes 6 and 7, we have
\begin{eqnarray}
\frac{\mathrm{d} (a_6^2/2)}{\mathrm{d} t}=
-\frac{a_{6}^2\,\left(\alpha ^2+\gamma ^2+\frac{5}{2}\right)}{\mathrm{Re}}
-\frac{3\,\sqrt{15}a_6\,a_{7}\,a_{11}\,\alpha \,\left(\alpha ^2-3\,\gamma ^2\right)}{14\,\left(\alpha ^2+\gamma ^2\right)}
+...
\end{eqnarray}
\begin{eqnarray}
\frac{\mathrm{d} (a_7^2/2)}{\mathrm{d} t}=
-\frac{a_{7}^2\,\left(\alpha ^2+\gamma ^2+\frac{5}{2}\right)}{\mathrm{Re}}+\frac{3\,\sqrt{15}\,a_{6}\,a_7\,a_{11}\,\alpha \,\left(\alpha ^2-3\,\gamma ^2\right)}{14\,\left(\alpha ^2+\gamma ^2\right)}+...
\end{eqnarray}
where only viscous term and the relevant coupling are shown for clarity; the correspondance between the coupling terms shows that there is an energy transfer between these two modes. Thus, the $L_z/2$ streak $a_{11}$ mediates energy exchanges between the two wall-normal vortices $a_6$ and $a_7$. As $a_6$ and $a_7$ are related to regeneration of $L_z$ and $L_z/2$ rolls, respectively, the non-linear terms involving these two wall-normal vortices couple the roll-streak structures at wavelengths $L_z$ and $L_z/2$.

\begin{figure}
\centerline{\includegraphics[width=0.5\textwidth]{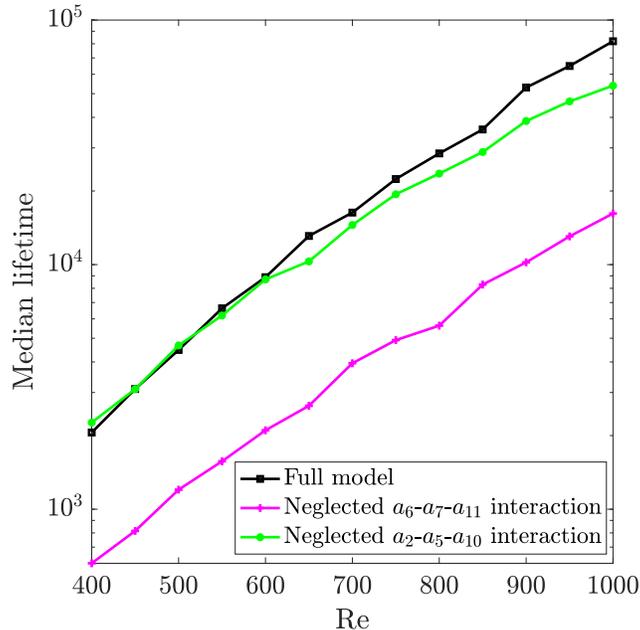}}
\caption{Median lifetimes of the model for Couette flow, neglecting the interaction between wall-normal vortices $a_6$ and $a_7$. Results from the full model, from fig. \ref{fig:couettelifetimes}(b), are repeated here for comparison.}
\label{fig:lifetimes_neglectQ}
\end{figure}

To confirm the dynamical relevance of the coupling between wall-normal vortices $a_6$ and $a_7$, we have obtained turbulence lifetimes for the Couette-flow mode artificially setting both $Q_{6,11,7}$ and $Q_{7,11,6}$ to zero in our model. By neglecting both terms the Galerkin model maintains the conservative nature of the quadratic term. Following the same procedure of the previous section, we have simulated 1000 initial conditions to compute turbulence lifetimes of the model neglecting this specific interaction between $a_6$ and $a_7$. The resulting median lifetimes are shown in fig. \ref{fig:lifetimes_neglectQ}, and compared to the results from the full model, repeated from fig. \ref{fig:couettelifetimes}(b). It is remarkable that neglecting only one energy exchange in the model leads to a reduction of turbulence lifetimes of almost an order of magnitude. Such results confirm that the relationship between rolls and streaks with different lengthscales is an important interaction maintaining turbulent motion for longer lifetimes.

\modif{An examination of the role of all non-linear interactions in the model is a complex task, and it is not straightforward to isolate the most relevant interactions in the dynamics. The results in figure \ref{fig:lifetimes_neglectQ} simply point out that $a_6$-$a_7$ interaction, mediated by $a_{11}$, is important for the observed long turbulence lifetimes. Other interactions are expected to be relevant as well in maintaining chaotic dynamics. However, some non-linear terms may have comparably lower influence in lifetimes. An example is also shown in figure \ref{fig:lifetimes_neglectQ}, with non-linear terms involving a triadic interaction among modes $a_2$, $a_5$ and $a_{10}$ removed from the model. For lower $\mathrm{Re}$ the impact on lifetimes is practically zero, and for higher $\mathrm{Re}$ there is a reduction of lifetimes once such interaction is neglected. However, the effect is much less significant than the order-of-magnitude reduction in median lifetimes once the $a_6$-$a_7$-$a_{11}$ interaction is discarded from the model. Thus, not all interactions are equally relevant in the longer turbulence lifetimes observed in the model, and the \modif{structure} interaction between $L_z$ and $L_z/2$ rolls and streaks, promoted by the $a_6$-$a_7$-$a_{11}$ triad, is here seen as particularly important.}

\section{Conclusions}
\label{sec:conclusions}

In this work a reduced-order model (ROM) for sinusoidal shear flow between parallel walls with free-slip boundary conditions (referred to as Waleffe flow) was derived using a Galerkin projection over Fourier modes, which are a natural basis for the velocity field. A larger basis including hundreds of modes was truncated to 12 modes by the requirement of a small Galerkin system leading to long transients of chaotic behaviour, preserving nonetheless the linear stability of the laminar solution for all Reynolds numbers. This led to a system of 12 ordinary differential equations. The same modes were then adapted to model Couette flow by rewriting the Galerkin system to velocity fluctuations, considering modes that are polynomials in the wall-normal direction in order to satisfy non-slip boundary conditions on the walls. Both Waleffe- and Couette-flow models considered periodicity over streamwise ($x$) and spanwise ($z)$ directions, which defines a computational box with respective lengths of $L_x$ and $L_z$. The retained modes included structures present in previous models~\citep{waleffe1997self,moehlis2004low}, but an important feature is the inclusion of two roll-streak structures, with spanwise wavelengths equal to $L_z$ and $L_z/2$.

The resulting models were explored considering $L_x=4\pi$ and $L_z=2\pi$. For such small computational domains it is known that Waleffe and Couette flow only display turbulent transients before returning to the laminar solution \citep{tuckerman2020patterns}, but the models in the present work lead to turbulence lifetimes that are orders of magnitude larger than similar models in the literature~\citep{eckhardt1999transition,moehlis2004low}. A critical amplitude threshold for the transition to turbulence scaling with $Re^{-2}$ was found, in agreement with the model of \citet{dawes2011turbulent} for Waleffe flow, which includes a larger number of spanwise Fourier modes in a system with 8 partial differential equations. The ROM for Couette flow was compared to results of direct numerical simulations (DNS), and despite the severe truncation to 12 modes the ROM results agree reasonably with mean and RMS profiles from the DNS, and also display larger-scale structures consistent with observations from the DNS snapshots. This highlights  that the ROM is able to model the salient features from the full DNS.

An important property of the models is that neglecting either of the roll modes leads to considerably lower turbulence lifetimes, which are reduced by more than an order of magnitude compared to the full model. This shows that the co-existence of roll-streak structures at the two spanwise lengthscales allowed by the model, $L_z$ and $L_z/2$, is an important feature to maintain long-lived chaotic dynamics. The interactions between the two lengthscales are rather subtle, as the models do not show non-linear interactions that directly couple them. Apart from the clearer coupling via the mean flow, which in both cases lead to amplification of streaks by the lift-up effect, there is a more subtle coupling of the $L_z$ and $L_z/2$ rolls, where each of them receives energy in a process involving one of the two wall-normal vortex modes in the ROM. These two wall-normal vortices have a non-linear coupling, which once neglected, so as to remove this indirect interaction between the rolls, is shown to lead to considerably lower turbulence lifetimes in the model. This shows that the intricate interaction between rolls and streaks with different wavelengths is an important feature of wall-bounded turbulent flows that maintain chaotic dynamics despite the linear stability of the laminar solution. Thus, reducing the dominant dynamics to a single wavenumber, as usual in the analysis of minimal flow units in small computational domains, may lead to the neglect of relevant interactions. The observations from the present models shows that the absence of $L_z/2$ rolls and streaks in previous ROMs \citep{waleffe1997self,moehlis2004low} leads to a truncation of the dynamics that is too severe, leading to relatively short-lived turbulence.

The availability of the present models, in particular the ROM for Couette flow, opens new directions for data analysis. Modal decomposition of flow databases has become a relevant area of turbulence research, as reviewed by \citet{taira2017modal}. Recent works have extracted coherent structures from flow databases, using spectral proper orthogonal decomposition, and compared them to results of resolvent analysis~\citep{schmidt2018spectral,lesshafft2019resolvent,abreu2020spectral}. When applied to turbulent flows, resolvent analysis is based on a linearisation around the turbulent mean flow, considering the (unknown) non-linear terms as an external forcing~\citep{mckeon2010critical}. Extraction of such ``forcing'' from non-linear terms in the Navier-Stokes system leads to an exact recovery of the flow statistics~\cite{morra2021colour}, but as such terms result from interactions among a broad range of frequencies and wavenumbers, this makes it difficult to determine which interactions are relevant in a given flow. Minimal flow units help in that task, and \modif{non-linear interactions have recently been studied in the resolvent framework by \citet{bae2019nonlinear} and \citet{nogueira_2021}.} The dynamical systems for Waleffe and Couette flows derived here may help in this task, as the twelve modes form an orthonormal basis that allows a straightforward projection of data in an alternative approach of modal decomposition, based here on a ROM. Non-linear interactions in the ROM can then be identified in a database from numerical simulation. As some non-linearities were here seen to be crucial to maintain turbulence for longer times, a capability to disrupt such interactions, by proper control action, could bring back the system to the desired laminar state. The present models may thus be useful in the identification of dominant non-linear effects in turbulent flows with low Reynolds numbers, hopefully pointing to new directions to flow control.

\section*{Acknowledgments}

I would like to thank Petr\^onio Nogueira, Eduardo Martini, Peter Jordan and Daniel Edgington-Mitchell for their comments on an early version of this manuscript. This work was supported by FAPESP grant 2019/27655-3 and CNPq grant 310523/2017-6. A numerical implementation of the present reduced-order models is available by request to the author.


\appendix

\section{Equations of the reduced-order model for Couette flow}
\label{sec:equations}

For Couette flow, by considering eqs. (\ref{eq:evenmodes}) and (\ref{eq:oddmodes}) to construct modes that are subsequently normalised, the basis functions of table \ref{tab:modes} become 
\begin{subequations}
\begin{equation}
\mathbf{u}_1=
\left(\begin{array}{c} -\frac{\sqrt{2}\,\sqrt{3}\,\sqrt{35}\,y\,\left(y^2-1\right)}{4}\\ 0\\ 0 \end{array}\right)
\end{equation}
\begin{equation}
\mathbf{u}_2=
\left(\begin{array}{c} 0\\ 0\\ -\frac{\sqrt{15}\,\sin\left(\alpha \,x\right)\,\left(y^2-1\right)}{2} \end{array}\right)
\end{equation}
\begin{equation}
\mathbf{u}_3=
\left(\begin{array}{c} 0\\ 0\\ -\frac{\sqrt{3}\,\sqrt{35}\,y\,\cos\left(\alpha \,x\right)\,\left(y^2-1\right)}{2} \end{array}\right)
\end{equation}
\begin{equation}
\mathbf{u}_4=
\left(\begin{array}{c} \frac{\sqrt{15}\,\sin\left(\gamma \,z\right)\,\left(y^2-1\right)}{2}\\ 0\\ 0 \end{array}\right)
\end{equation}
\begin{equation}
\mathbf{u}_5=
\left(\begin{array}{c} 0\\ -\frac{3\,\sqrt{35}\,\gamma \,\sin\left(\gamma \,z\right)\,{\left(y^2-1\right)}^2}{8\,\sqrt{\gamma ^2+3}}\\ -\frac{3\,\sqrt{35}\,y\,\cos\left(\gamma \,z\right)\,\left(y^2-1\right)}{2\,\sqrt{\gamma ^2+3}} \end{array}\right)
\end{equation}
\begin{equation}
\mathbf{u}_6=
\left(\begin{array}{c} -\frac{\sqrt{30}\,\gamma \,\cos\left(\gamma \,z\right)\,\sin\left(\alpha \,x\right)\,\left(y^2-1\right)}{2\,\sqrt{\alpha ^2+\gamma ^2}}\\ 0\\ \frac{\sqrt{30}\,\alpha \,\cos\left(\alpha \,x\right)\,\sin\left(\gamma \,z\right)\,\left(y^2-1\right)}{2\,\sqrt{\alpha ^2+\gamma ^2}} \end{array}\right)
\end{equation}
\begin{equation}
\mathbf{u}_7=
\left(\begin{array}{c} \frac{\sqrt{30}\,\gamma \,\cos\left(\alpha \,x\right)\,\sin\left(\gamma \,z\right)\,\left(y^2-1\right)}{2\,\sqrt{\alpha ^2+\gamma ^2}}\\ 0\\ -\frac{\sqrt{30}\,\alpha \,\cos\left(\gamma \,z\right)\,\sin\left(\alpha \,x\right)\,\left(y^2-1\right)}{2\,\sqrt{\alpha ^2+\gamma ^2}} \end{array}\right)
\end{equation}
\begin{equation}
\mathbf{u}_8=
\left(\begin{array}{c} -\frac{3\,\sqrt{70}\,y\,\cos\left(\alpha \,x\right)\,\cos\left(\gamma \,z\right)\,\left(y^2-1\right)}{2\,\sqrt{\alpha ^2+3}}\\ -\frac{3\,\sqrt{70}\,\alpha \,\cos\left(\gamma \,z\right)\,\sin\left(\alpha \,x\right)\,{\left(y^2-1\right)}^2}{8\,\sqrt{\alpha ^2+3}}\\ 0 \end{array}\right)
\end{equation}
\begin{equation}
\mathbf{u}_9=
\left(\begin{array}{c} \frac{3\,\sqrt{70}\,y\,\sin\left(\alpha \,x\right)\,\sin\left(\gamma \,z\right)\,\left(y^2-1\right)}{2\,\sqrt{\alpha ^2+3}}\\ -\frac{3\,\sqrt{70}\,\alpha \,\cos\left(\alpha \,x\right)\,\sin\left(\gamma \,z\right)\,{\left(y^2-1\right)}^2}{8\,\sqrt{\alpha ^2+3}}\\ 0 \end{array}\right)
\end{equation}
\begin{equation}
\mathbf{u}_{10}=
\left(\begin{array}{c} \frac{\sqrt{210}\,\alpha \,\gamma \,y\,\cos\left(\alpha \,x\right)\,\cos\left(\gamma \,z\right)\,\left(y^2-1\right)}{2\,\sqrt{\alpha ^2+3}\,\sqrt{\alpha ^2+\gamma ^2+3}}\\ -\frac{3\,\sqrt{210}\,\gamma \,\cos\left(\gamma \,z\right)\,\sin\left(\alpha \,x\right)\,{\left(y^2-1\right)}^2}{8\,\sqrt{\alpha ^2+3}\,\sqrt{\alpha ^2+\gamma ^2+3}}\\ \frac{\sqrt{210}\,y\,\sin\left(\alpha \,x\right)\,\sin\left(\gamma \,z\right)\,\left(y^2-1\right)\,\left(2\,\alpha ^2+6\right)}{4\,\sqrt{\alpha ^2+3}\,\sqrt{\alpha ^2+\gamma ^2+3}} \end{array}\right)
\end{equation}
\begin{equation}
\mathbf{u}_{11}=
\left(\begin{array}{c} \frac{\sqrt{15}\,\sin\left(2\,\gamma \,z\right)\,\left(y^2-1\right)}{2}\\ 0\\ 0 \end{array}\right)
\end{equation}
\begin{equation}
\mathbf{u}_{12}=
\left(\begin{array}{c} 0\\ -\frac{3\,\sqrt{35}\,\gamma \,\sin\left(2\,\gamma \,z\right)\,{\left(y^2-1\right)}^2}{4\,\sqrt{4\,\gamma ^2+3}}\\ -\frac{3\,\sqrt{35}\,y\,\cos\left(2\,\gamma \,z\right)\,\left(y^2-1\right)}{2\,\sqrt{4\,\gamma ^2+3}} \end{array}\right).
\end{equation}
\label{eq:couettemodes}
\end{subequations}

Auxiliary wavenumbers are defined as for Waleffe flow, considering $\beta = \sqrt{3}$. The resulting Galerkin system for fluctuations around the laminar solution is
\begin{subequations}
\begin{eqnarray}
\frac{\mathrm{d} a_1}{\mathrm{d} t}=
-\frac{21\,a_{1}}{2\,\mathrm{Re}} + \frac{3\,\sqrt{10}\,a_{4}\,a_{5}\,\gamma }{4\,k_{\beta,\gamma}}+\frac{3\,\sqrt{10}\,a_{11}\,a_{12}\,\gamma }{2\,k_{\beta,2\gamma}}
-\frac{3\,\sqrt{10}\,a_{6}\,a_{8}\,\alpha \,\gamma }{4\,k_{\alpha,\beta}\,k_{\alpha,\gamma}}\nonumber \\
+\frac{3\,\sqrt{10}\,a_{7}\,a_{9}\,\alpha \,\gamma }{4\,k_{\alpha,\beta}\,k_{\alpha,\gamma}}-\frac{3\,\sqrt{30}\,a_{6}\,a_{10}\,\gamma ^2}{4\,k_{\alpha,\beta}\,k_{\alpha,\gamma}\,k_{\alpha,\beta,\gamma}}
\label{eq:a1couette}
\end{eqnarray}
\begin{eqnarray}
\frac{\mathrm{d} a_2}{\mathrm{d} t}=
-\frac{a_{2}\,\alpha ^2+\frac{5\,a_{2}}{2}}{\mathrm{Re}}+\frac{\sqrt{7}\,a_{3}\,\alpha }{7}+\frac{\sqrt{30}\,a_{1}\,a_{3}\,\alpha }{6}+\frac{3\,\sqrt{30}\,a_{4}\,a_{6}\,\alpha ^2}{14\,k_{\alpha,\gamma}}\nonumber \\
+\frac{3\,\sqrt{30}\,a_{5}\,a_{8}\,\alpha }{4\,k_{\alpha,\beta}\,k_{\beta,\gamma}}-\frac{3\,\sqrt{10}\,a_{5}\,a_{10}\,\alpha ^2\,\gamma }{4\,k_{\alpha,\beta}\,k_{\beta,\gamma}\,k_{\alpha,\beta,\gamma}}
\end{eqnarray}
\begin{eqnarray}
\frac{\mathrm{d} a_3}{\mathrm{d} t}=
-\frac{a_{3}\,\left(\alpha ^2+\frac{21}{2}\right)}{\mathrm{Re}}+\frac{\sqrt{10}\,a_{5}\,a_{6}\,\alpha \,\gamma }{4\,k_{\beta,\gamma}\,k_{\alpha,\gamma}}\nonumber \\
-\frac{\sqrt{30}\,a_{1}\,a_{2}\,\alpha }{6}-\frac{\sqrt{30}\,a_{4}\,a_{10}\,\alpha \,k_{\alpha,\beta}}{6\,k_{\alpha,\beta,\gamma}}-\frac{\sqrt{7}\,a_{2}\,\alpha }{7}
\end{eqnarray}
\begin{eqnarray}
\frac{\mathrm{d} a_4}{\mathrm{d} t}=
-\frac{a_{4}\,\left(\gamma ^2+\frac{5}{2}\right)}{\mathrm{Re}}
-\frac{3\,\sqrt{21}\,a_{5}\,\gamma }{14\,k_{\beta,\gamma}}-\frac{3\,\sqrt{10}\,a_{1}\,a_{5}\,\gamma }{4\,k_{\beta,\gamma}}
+\frac{\sqrt{30}\,a_{3}\,a_{10}\,\alpha \,\gamma ^2}{6\,k_{\alpha,\beta}\,k_{\alpha,\beta,\gamma}}\nonumber \\
-\frac{\sqrt{10}\,a_{3}\,a_{8}\,\gamma }{2\,k_{\alpha,\beta}}
-\frac{3\,\sqrt{30}\,a_{2}\,a_{6}\,\gamma ^2}{14\,k_{\alpha,\gamma}}
\end{eqnarray}
\begin{eqnarray}
\frac{\mathrm{d} a_5}{\mathrm{d} t}=
-\frac{a_{5}\,\left(2\,\gamma ^4+12\,\gamma ^2+63\right)}{2\,\mathrm{Re}\,\left(\gamma ^2+3\right)}+\frac{\sqrt{10}\,a_{3}\,a_{6}\,\alpha \,\gamma }{k_{\beta,\gamma}\,k_{\alpha,\gamma}}+\frac{\sqrt{30}\,a_{2}\,a_{8}\,\alpha \,\left(10\,\gamma ^2-33\right)}{44\,k_{\alpha,\beta}\,k_{\beta,\gamma}}\nonumber \\
+\frac{\sqrt{10}\,a_{2}\,a_{10}\,\gamma \,\left(44\,\alpha ^2+30\,\gamma ^2+33\right)}{44\,k_{\alpha,\beta}\,k_{\beta,\gamma}\,k_{\alpha,\beta,\gamma}}
\end{eqnarray}
\begin{eqnarray}
\frac{\mathrm{d} a_6}{\mathrm{d} t}=
-\frac{a_{6}\,\left(\alpha ^2+\gamma ^2+\frac{5}{2}\right)}{\mathrm{Re}}
+\frac{5\,\sqrt{21}\,a_{8}\,\alpha \,\gamma }{14\,k_{\alpha,\beta}\,k_{\alpha,\gamma}}-\frac{3\,\sqrt{30}\,a_{2}\,a_{4}\,\left(\alpha ^2-\gamma ^2\right)}{14\,k_{\alpha,\gamma}}
\nonumber \\
-\frac{3\,\sqrt{15}\,a_{7}\,a_{11}\,\alpha \,\left(\alpha ^2-3\,\gamma ^2\right)}{14\,\left(\alpha ^2+\gamma ^2\right)}
-\frac{\sqrt{7}\,a_{10}\,\left(2\,\alpha ^4+2\,\alpha ^2\,\gamma ^2+6\,\alpha ^2-9\,\gamma ^2\right)}{14\,k_{\alpha,\beta}\,k_{\alpha,\gamma}\,k_{\alpha,\beta,\gamma}}
\nonumber \\
+\frac{5\,\sqrt{10}\,a_{1}\,a_{8}\,\alpha \,\gamma }{4\,k_{\alpha,\beta}\,k_{\alpha,\gamma}}
-\frac{5\,\sqrt{10}\,a_{3}\,a_{5}\,\alpha \,\gamma }{4\,k_{\beta,\gamma}\,k_{\alpha,\gamma}}
\nonumber \\
-\frac{\sqrt{30}\,a_{1}\,a_{10}\,\left(2\,\alpha ^4+2\,\alpha ^2\,\gamma ^2+6\,\alpha ^2-9\,\gamma ^2\right)}{12\,k_{\alpha,\beta}\,k_{\alpha,\gamma}\,k_{\alpha,\beta,\gamma}}+\frac{\sqrt{15}\,a_{9}\,a_{12}\,\left(3\,\alpha ^2-4\,\gamma ^2\right)}{4\,k_{\alpha,\beta}\,k_{\beta,2\gamma}\,k_{\alpha,\gamma}}
\end{eqnarray}
\begin{eqnarray}
\frac{\mathrm{d} a_7}{\mathrm{d} t}=
-\frac{a_{7}\,\left(\alpha ^2+\gamma ^2+\frac{5}{2}\right)}{\mathrm{Re}}+\frac{3\,\sqrt{15}\,a_{6}\,a_{11}\,\alpha \,\left(\alpha ^2-3\,\gamma ^2\right)}{14\,\left(\alpha ^2+\gamma ^2\right)}\nonumber \\
-\frac{5\,\sqrt{21}\,a_{9}\,\alpha \,\gamma }{14\,k_{\alpha,\beta}\,k_{\alpha,\gamma}}
-\frac{5\,\sqrt{10}\,a_{1}\,a_{9}\,\alpha \,\gamma }{4\,k_{\alpha,\beta}\,k_{\alpha,\gamma}}+\frac{\sqrt{15}\,a_{8}\,a_{12}\,\left(3\,\alpha ^2-4\,\gamma ^2\right)}{4\,k_{\alpha,\beta}\,k_{\beta,2\gamma}\,k_{\alpha,\gamma}}\nonumber \\
-\frac{\sqrt{5}\,a_{10}\,a_{12}\,\alpha \,\gamma \,\left(8\,\alpha ^2-4\,\gamma ^2+15\right)}{4\,k_{\alpha,\beta}\,k_{\beta,2\gamma}\,k_{\alpha,\gamma}\,k_{\alpha,\beta,\gamma}}
\label{eq:a7couette}
\end{eqnarray}
\begin{eqnarray}
\frac{\mathrm{d} a_8}{\mathrm{d} t}=
-\frac{\frac{a_{8}\,\left(2\,\alpha ^4+2\,\alpha ^2\,\gamma ^2+12\,\alpha ^2+6\,\gamma ^2+63\right)}{2\,\left(\alpha ^2+3\right)}-\frac{15\,\sqrt{3}\,a_{10}\,\alpha \,\gamma }{2\,\left(\alpha ^2+3\right)\,k_{\alpha,\beta,\gamma}}}{\mathrm{Re}} + \frac{\sqrt{10}\,a_{3}\,a_{4}\,\gamma }{2\,k_{\alpha,\beta}}\nonumber \\
-\frac{\sqrt{21}\,a_{6}\,\alpha \,\gamma }{7\,k_{\alpha,\beta}\,k_{\alpha,\gamma}}-\frac{\sqrt{15}\,a_{9}\,a_{11}\,\alpha \,\left(10\,\alpha ^2+11\right)}{44\,\left(\alpha ^2+3\right)}-\frac{\sqrt{10}\,a_{1}\,a_{6}\,\alpha \,\gamma }{2\,k_{\alpha,\beta}\,k_{\alpha,\gamma}}\nonumber \\
-\frac{5\,\sqrt{30}\,a_{2}\,a_{5}\,\alpha \,\gamma ^2}{22\,k_{\alpha,\beta}\,k_{\beta,\gamma}}
-\frac{\sqrt{15}\,a_{7}\,a_{12}\,\gamma ^2\,\left(10\,\alpha ^2-11\right)}{11\,k_{\alpha,\beta}\,k_{\beta,2\gamma}\,k_{\alpha,\gamma}}
\end{eqnarray}
\begin{eqnarray}
\frac{\mathrm{d} a_9}{\mathrm{d} t}=
\frac{\sqrt{21}\,a_{7}\,\alpha \,\gamma }{7\,k_{\alpha,\beta}\,k_{\alpha,\gamma}}-\frac{a_{9}\,\left(2\,\alpha ^4+2\,\alpha ^2\,\gamma ^2+12\,\alpha ^2+6\,\gamma ^2+63\right)}{2\,\mathrm{Re}\,\left(\alpha ^2+3\right)}\nonumber \\
+\frac{\sqrt{15}\,a_{8}\,a_{11}\,\alpha \,\left(10\,\alpha ^2+11\right)}{44\,\left(\alpha ^2+3\right)}+\frac{\sqrt{10}\,a_{1}\,a_{7}\,\alpha \,\gamma }{2\,k_{\alpha,\beta}\,k_{\alpha,\gamma}}\nonumber \\
-\frac{3\,\sqrt{5}\,a_{10}\,a_{11}\,\gamma \,\left(12\,\alpha ^2+55\right)}{44\,\left(\alpha ^2+3\right)\,k_{\alpha,\beta,\gamma}}-\frac{\sqrt{15}\,a_{6}\,a_{12}\,\gamma ^2\,\left(10\,\alpha ^2-11\right)}{11\,k_{\alpha,\beta}\,k_{\beta,2\gamma}\,k_{\alpha,\gamma}}
\end{eqnarray}
\begin{eqnarray}
\frac{\mathrm{d} a_{10}}{\mathrm{d} t}=
-\frac{a_{10}\, \kappa }{2\,\left(\alpha ^2+3\right)\,\left(\alpha ^2+\gamma ^2+3\right)\mathrm{Re}} 
-\frac{15\,\sqrt{3}\,a_{8}\,\alpha \,\gamma }{2\,\left(\alpha ^2+3\right)\,k_{\alpha,\beta,\gamma}\mathrm{Re}}\nonumber \\
+\frac{\sqrt{7}\,a_{6}\,\alpha ^2\,k_{\alpha,\beta,\gamma}}{7\,k_{\alpha,\beta}\,k_{\alpha,\gamma}}+\frac{\sqrt{30}\,a_{1}\,a_{6}\,\alpha ^2\,k_{\alpha,\beta,\gamma}}{6\,k_{\alpha,\beta}\,k_{\alpha,\gamma}}
-\frac{19\,\sqrt{5}\,a_{9}\,a_{11}\,\alpha ^2\,\gamma }{44\,\left(\alpha ^2+3\right)\,k_{\alpha,\beta,\gamma}}\nonumber \\
+\frac{\sqrt{30}\,a_{3}\,a_{4}\,\alpha \,\left(\alpha ^2-\gamma ^2+3\right)}{6\,k_{\alpha,\beta}\,k_{\alpha,\beta,\gamma}}
-\frac{\sqrt{10}\,a_{2}\,a_{5}\,\gamma \,\left(11\,\alpha ^2+30\,\gamma ^2+33\right)}{44\,k_{\alpha,\beta}\,k_{\beta,\gamma}\,k_{\alpha,\beta,\gamma}}\nonumber \\
-\frac{41\,\sqrt{5}\,a_{7}\,a_{12}\,\alpha \,\gamma ^3}{11\,k_{\alpha,\beta}\,k_{\beta,2\gamma}\,k_{\alpha,\gamma}\,k_{\alpha,\beta,\gamma}}
\end{eqnarray}
\begin{eqnarray}
\frac{\mathrm{d} a_{11}}{\mathrm{d} t}=
-\frac{a_{11}\,\left(4\,\gamma ^2+\frac{5}{2}\right)}{\mathrm{Re}}+
\frac{5\,\sqrt{5}\,a_{9}\,a_{10}\,\gamma }{4\,k_{\alpha,\beta,\gamma}}-\frac{3\,\sqrt{21}\,a_{12}\,\gamma }{7\,k_{\beta,2\gamma}}-\frac{3\,\sqrt{10}\,a_{1}\,a_{12}\,\gamma }{2\,k_{\beta,2\gamma}}
\end{eqnarray}
\begin{eqnarray}
\frac{\mathrm{d} a_{12}}{\mathrm{d} t}=
-\frac{a_{12}\,\left(32\,\gamma ^4+48\,\gamma ^2+63\right)}{2\,\mathrm{Re}\,\left(4\,\gamma ^2+3\right)}+
\frac{\sqrt{15}\,a_{6}\,a_{9}\,\alpha ^2\,\left(40\,\gamma ^2-33\right)}{44\,k_{\alpha,\beta}\,k_{\beta,2\gamma}\,k_{\alpha,\gamma}}\nonumber \\
+\frac{\sqrt{15}\,a_{7}\,a_{8}\,\alpha ^2\,\left(40\,\gamma ^2-33\right)}{44\,k_{\alpha,\beta}\,k_{\beta,2\gamma}\,k_{\alpha,\gamma}}+\frac{\sqrt{5}\,a_{7}\,a_{10}\,\alpha \,\gamma \,\left(88\,\alpha ^2+120\,\gamma ^2+165\right)}{44\,k_{\alpha,\beta}\,k_{\beta,2\gamma}\,k_{\alpha,\gamma}\,k_{\alpha,\beta,\gamma}}
\end{eqnarray}
\label{eq:couettemodel}
\end{subequations}
where $\kappa = \left(2\,\alpha ^6+4\,\alpha ^4\,\gamma ^2+33\,\alpha ^4+2\,\alpha ^2\,\gamma ^4+39\,\alpha ^2\,\gamma ^2+144\,\alpha ^2+6\,\gamma ^4+36\,\gamma ^2+189\right)$.

\section{Results for other computational domains}
\label{sec:boxsize}

\begin{figure}
\centerline{\includegraphics[width=0.5\textwidth]{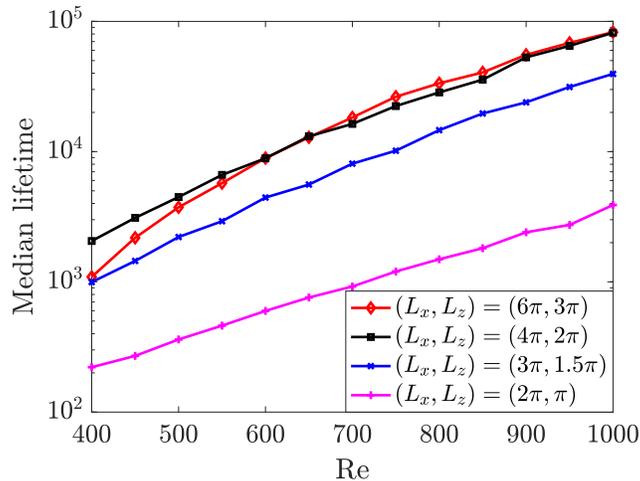}}
\caption{Effect of computational domain on turbulence lifetimes. Results from fig. \ref{fig:couettelifetimes} are compared to median lifetimes with other domain sizes, calculated from 1000 simulations starting from random initial condition with norm equal to 0.3.}
\label{fig:boxsize}
\end{figure}

The analysis in this work used a computational domain with $L_x=4\pi$ and $L_z=2\pi$, but the conclusions do not depend on this particular choice. Figure \ref{fig:boxsize} shows median turbulence lifetimes for two other domain sizes, compared to the reference results for $L_x=4\pi$ and $L_z=2\pi$. The results show that a larger domain, with $L_x=6\pi$ and $L_z=3\pi$, leads to very similar lifetimes, except for the lower Reynolds numbers considered, whereas the smaller domain, with $L_x=3\pi$ and $L_z=1.5\pi$, has lower lifetimes by a factor of about 2. If the domain size is further decreased to $L_x=2\pi$ and $L_z=1\pi$ turbulence lifetimes are more affected, with values about an order of magnitude lower than what is found for the larger domains. The lifetime of 220 for $\mathrm{Re}=400$ is of the same order of the lifetime of 400 by \citet{kreilos2014increasing} using DNS.

Together with the analysis in this work, these results indicate that box sizes should be large enough to accommodate both $L_z$ and $L_z/2$ rolls and streaks. Small computational domains would lead to low $L_z/2$ wavelengths, which would lead to stronger damping of the $L_z/2$ rolls and streaks.



\end{document}